# High sensitivity Cavity Ring Down spectroscopy of $^{18}O$ enriched carbon dioxide between 5850 and 7000 cm$^{-1}$: III. Analysis and theoretical modelling of the $^{12}C^{17}O_2$, $^{16}O^{12}C^{17}O$, $^{17}O^{12}C^{18}O$, $^{16}O^{13}C^{17}O$ and $^{17}O^{13}C^{18}O$ spectra.

E. V. Karlovets [ab], A. Campargue [a*], D. Mondelain [a], S. Kassi [a], S. A. Tashkun [b], V. I. Perevalov [b]

[a] *Université Grenoble 1/CNRS, UMR5588 LIPhy, Grenoble, F-38041, France*
[b]*Laboratory of Theoretical Spectroscopy, V.E. Zuev Institute of Atmospheric Optics, SB, Russian Academy of Science, 1, Akademician Zuev square, 634021, Tomsk, Russia*

Number of pages: 35

Number of tables: 15

Number of figures: 10

* Corresponding author: Alain.Campargue@ujf-grenoble.fr

## *Abstract*

More than 19700 transitions belonging to eleven isotopologues of carbon dioxide have been assigned in the room temperature absorption spectrum of highly $^{18}O$ enriched carbon dioxide recorded by very high sensitivity CW-Cavity Ring Down spectroscopy between 5851 and 6990 cm$^{-1}$ (1.71-1.43 µm). This third and last report is devoted to the analysis of the bands of five $^{17}O$ containing isotopologues present at very low concentration in the studied spectra: $^{16}O^{12}C^{17}O$, $^{17}O^{12}C^{18}O$, $^{16}O^{13}C^{17}O$, $^{17}O^{13}C^{18}O$ and $^{12}C^{17}O_2$ (627, 728, 637, 738 and 727 in short hand notation). On the basis of the predictions of effective Hamiltonian models, a total of 1759, 1786, 335, 273 and 551 transitions belonging to 24, 24, 5, 4 and 7 bands were rovibrationnally assigned for 627, 728, 637, 738 and 727, respectively. For comparison, only five bands were previously measured in the region for the 728 species. All the identified bands belong to the $\Delta P$= 8 and 9 series of transitions, where $P = 2V_1 + V_2 + 3V_3$ is the polyad number ($V_i$ are vibrational quantum numbers). The band-by-band analysis has allowed deriving accurate spectroscopic parameters of 61 bands from a fit of the measured line positions. Two interpolyad resonance perturbations were identified.

Using the newly measured line positions and those collected from the literature, the global modeling of the line positions within the effective Hamiltonian approach was performed and a new set of Hamiltonian parameters was obtained for the $^{17}O^{12}C^{18}O$, $^{16}O^{13}C^{17}O$ and $^{17}O^{13}C^{18}O$ isotopologues. For the five studied isotopologues, the effective dipole moment parameters of the $\Delta P$= 8 and 9 series were derived from a global fit of the measured line intensities. The obtained results will be used for the improvement of the quality of the line positions and intensities in the most currently used spectroscopic databases of carbon dioxide.

## 1. Introduction

This contribution is the third and last one devoted to the analysis of the absorption spectrum of $^{18}O$ enriched carbon dioxide highly that we recorded by very high sensitivity CW-Cavity Ring Down spectroscopy (CW-CRDS) between 5851 and 6990 $cm^{-1}$ (1.71-1.43 µm) [1,2]. In a first report, the analysis and modeling of the $^{16}O^{12}C^{18}O$ line positions and line intensities were published [1]. The second contribution was devoted to three multiply substituted isotopologues: $^{12}C^{18}O_2$, $^{13}C^{18}O_2$ and $^{16}O^{13}C^{18}O$ [2]. This paper deals with the analysis and theoretical modelling of the spectra of $^{17}O$ containing isotopologues: $^{12}C^{17}O_2$, $^{16}O^{12}C^{17}O$, $^{17}O^{12}C^{18}O$, $^{16}O^{13}C^{17}O$ and $^{17}O^{13}C^{18}O$. Indeed, the used sample presented not only high $^{18}O$ enrichment (more than 50 % of the oxygen atoms) but also a significant enrichment in $^{17}O$ (about 2 %). Although very small (between $2\times10^{-4}$ and $2\times10^{-2}$), the relative concentrations of the five studied $^{17}O$ isotopologues were significantly higher than their natural values. This $^{17}O$ enrichment combined with the very high sensitivity of the recorded spectra (routine sensitivity better than $1\times10^{-10}$ $cm^{-1}$) has allowed for the detection of a large amount of new bands for the five above listed $^{17}O$ isotopologues. This last contribution is devoted to their assignment and analysis.

## 2. Experiment

Our fibered CW-CRDS spectrometer is described in detail in Ref. [3] while the data acquisition has been presented in Ref. [1]. The routine sensitivity of the recordings ranges between $4\times10^{-11}$ and $1\times10^{-10}$ $cm^{-1}$ as illustrated on Fig. 1.

Two series of spectra were recorded for pressure values of 0.20 and 5.0 Torr. The temperature of the different recordings was observed to vary by ±0.5 K around an average value of 295.9 K. The $CO_2$ sample (from Sigma-Aldrich) has a stated chemical purity better than 99.9 % while the stated relative abundance of the oxygen atoms is: $^{16}O$: 47.3%, $^{17}O$: 2.0% and $^{18}O$: 50.7%. The relative abundances of the five isotopologues studied in this paper are presented in Table 1. They were obtained assuming a statistical distribution of the O atoms in the various $CO_2$ isotopologues. The validity of this assumption can be tested using the intensities of the 2925 $^{12}C^{16}O_2$ lines and 662 $^{13}C^{16}O_2$ lines identified in the spectrum. Using their corresponding intensities provided in the HITRAN database, we found that the $^{12}C^{16}O_2$ relative abundance agrees within 7 % with its calculated statistical value (22.15 %). The $^{13}C^{16}O_2$ relative abundance was found in excellentagreement (deviation less than 1 %) with its statistical value based on a “natural” $^{13}C/^{12}C$ relative abundance.

The spectra were calibrated with the help of a wavelength meter and highly accurate reference positions of $CO_2$ and $H_2O$ lines (present as impurities) as provided by the HITRAN database [4]. The uncertainty of the wavenumber calibration is believed to be less than 1× $10^{-3}$ $cm^{-1}$. The line centers and intensities were determined using an interactive least squares multi-line fitting program assuming a Voigt function for the line profile (see Ref. [1] for details).

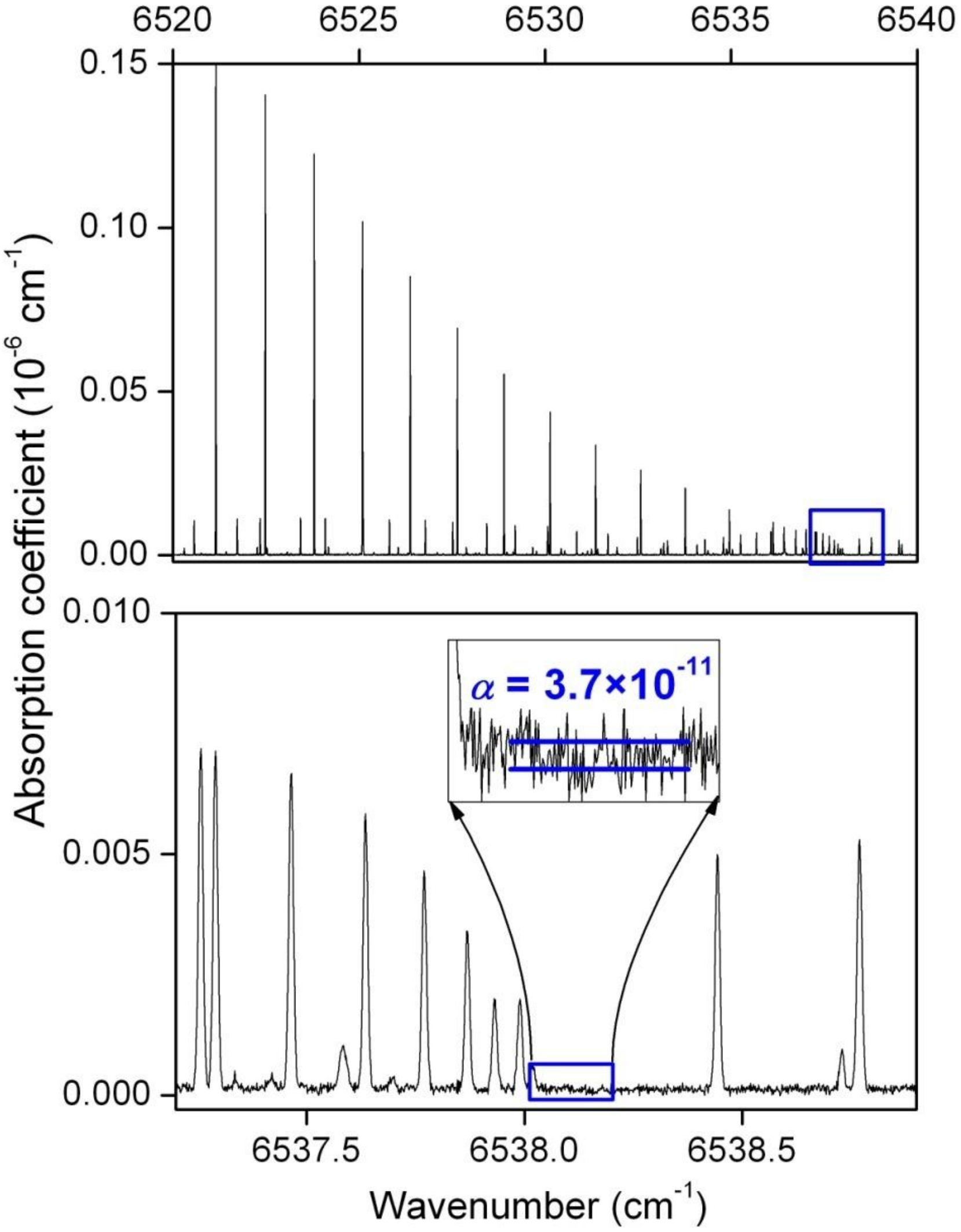

**Fig. 1.** CRDS spectrum of carbon dioxide enriched in $^{18}O$ at a pressure of 0.20 Torr. The *Q* branch of the 11122-00001 band of $^{12}C^{16}O_2$ at 6537.102 $cm^{-1}$ is observed superimposed to the newly observed strong *R* branch of the 11122-00001 band of $^{16}O^{12}C^{17}O$ at 6505.30 $cm^{-1}$.

**Table 1.** Relative isotopic abundances of the five $^{17}O$ isotopologues of $CO_2$ in the studied sample

| Isotopologue | Abundance (in %) |
|---|---|
| $^{12}C^{17}O_2$ (727) | 0.040 |
| $^{16}O^{12}C^{17}O$ (627) | 1.87 |
| $^{17}O^{12}C^{18}O$ (728) | 2.01 |
| $^{16}O^{13}C^{17}O$ (637) | 0.020 |
| $^{17}O^{13}C^{18}O$ (738) | 0.020 |

## 3. Rovibrational assignment

The high dynamics on the intensity scale provided by the CRDS technique and the used of two pressure samples allowed detecting lines with intensities ranging over five orders of magnitude. As a result, the resulting global line list includes more than 20700 entries (about 17 lines/cm$^{-1}$) which made the assignment particularly laborious. The spectral congestion illustrated in Fig. 2, is due to the superposition of lines of eleven of the twelve stable $CO_2$ isotopologues (only $^{13}C^{17}O_2$ lines could not be identified) and trace gas species present as impurities in the sample. By comparison with Refs. [4,5], lines of the various oxygen isotopologues of CO and $H_2O$,of methane (near 6000 cm$^{-1}$) and even 15 lines of hydrogen cyanide (near 6500 cm$^{-1}$) were identified. The estimated abundance values of the identified impurities are given in Table 2. Their overall concentration is on the order of 0.1 %.

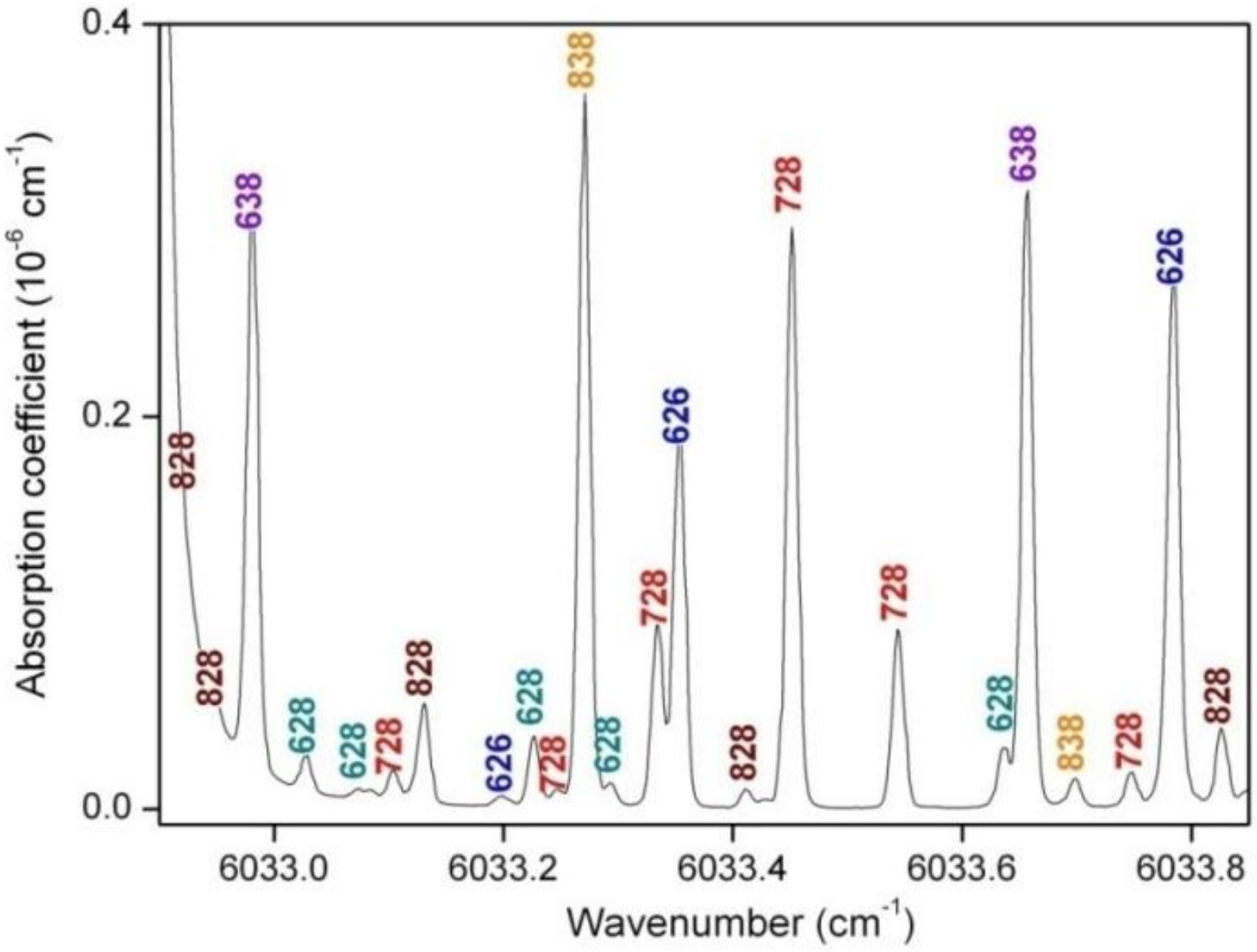

**Fig. 2.** CW-CRDS spectrum of $^{18}O$ enriched carbon dioxide recorded at 5 Torr near 6033 cm$^{-1}$. In the 1 cm$^{-1}$ wide displayed interval, lines due to six isotopologues are assigned.

The assignments were carried out on the basis of the prediction of the global effective Hamiltonian (EH) model [6,7,8]. The parameters of the effective Hamiltonians for the $^{16}O^{12}C^{17}O$ and $^{16}O^{13}C^{17}O$ isotopologues were taken from Ref. [9] and Ref. [10], respectively, and those for the $^{12}C^{17}O_2$, $^{17}O^{12}C^{18}O$ and $^{17}O^{13}C^{18}O$ isotopologues were obtained using isotopic extrapolation method presented in Ref. [11]. Line intensity predictions were also crucial in the assignment process. In the case of the $\Delta P$= 9 series of transitions, line intensities were calculated using the respective effective dipole moment (EDM) parameters of the principal isotopologue [12] whereas in the case of the $\Delta P$= 8 series of $^{16}O^{12}C^{17}O$ and $^{17}O^{12}C^{18}O$, the calculated relative intensities were used.

**Table 2.** Impurities identified in the studied spectrum between 5851 and 6990 cm$^{-1}$ with corresponding number of lines and relative concentrations.

| Impurities | Number of observed transitions | Relative abundance (×10$^{-4}$) |
|---|---|---|
| Carbon monoxide | | |
| $^{12}C^{16}O$ | 23 | 1.5 |
| $^{12}C^{17}O$ | 15 | 0.25 |
| $^{12}C^{18}O$ | 32 | 6.9 |
| Water | | |
| $H_2{}^{16}O$ | 789 | 0.38 |
| $H_2{}^{18}O$ | 200 | 3.4 |
| $H_2{}^{17}O$ | 29 | 0.35 |
| $HD^{16}O$ | 217 | 0.65 |
| Methane | | |
| $^{12}CH_4$ | 26 | 1.6 |
| $^{12}CH_3D$ | 22 | 0.9 |

**Table 3.** Comparison of the number of transitions and bands of $^{12}C^{17}O_2$, $^{16}O^{12}C^{17}O$, $^{17}O^{12}C^{18}O$, $^{16}O^{13}C^{17}O$ and $^{17}O^{13}C^{18}O$ reported in the literature and obtained in this work in the 5851-6990 cm$^{-1}$ region.

| Isotopologue | Previous/This work | Technique/enrichment | Number of transitions | Number of bands |
|---|---|---|---|---|
| **$^{16}O^{12}C^{17}O$** | Lyulin et al. [13] | FTS/ $^{17}O$ and $^{18}O$ | 731 | 8 |
| | Perevalov et al.[14] | CW-CRDS/ natural | 767 | 11 |
| | **This work** | **CW-CRDS/ $^{18}O$** | **1759** | **24** |
| **$^{17}O^{12}C^{18}O$** | X. de Ghellinck d'Elseghem Vaernewijck et al. [15,16] | OPO–Femto-FT-CEAS/ $^{17}O$ | 261 | 4 |
| | Lyulin et al. [13] | FTS/ $^{17}O$ and $^{18}O$ | 314 | 4 |
| | **This work** | **CW-CRDS/ $^{18}O$** | **1786** | **24** |
| **$^{16}O^{13}C^{17}O$** | Perevalov et al.[10] | CW-CRDS/ natural and $^{13}C$ | 115 | 2 |
| | Perevalov et al.[17] | CW-CRDS/ $^{13}C$ | 404 | 11 |
| | Ding et al.[18] | CW-CRDS/ $^{13}C$ | 266 | 3 |
| | Perevalov et al.[19] | CW-CRDS/natural and $^{13}C$ | 249 | 3 |
| | **This work** | **CW-CRDS/ $^{18}O$** | **335** | **5** |
| **$^{17}O^{13}C^{18}O$** | Perevalov et al.[17] | CW-CRDS/ $^{13}C$ | 130 | 3 |
| | **This work** | **CW-CRDS/ $^{18}O$** | **273** | **4** |
| **$^{12}C^{17}O_2$** | Lyulin et al. [13] | FTS/ $^{17}O$ and $^{18}O$ | 621 | 7 |
| | X. de Ghellinck d'Elseghem Vaernewijck et al. [15,16] | OPO–Femto-FT-CEAS/ $^{17}O$ | 398 | 4 |
| | **This work** | **CW-CRDS/ $^{18}O$** | **551** | **7** |

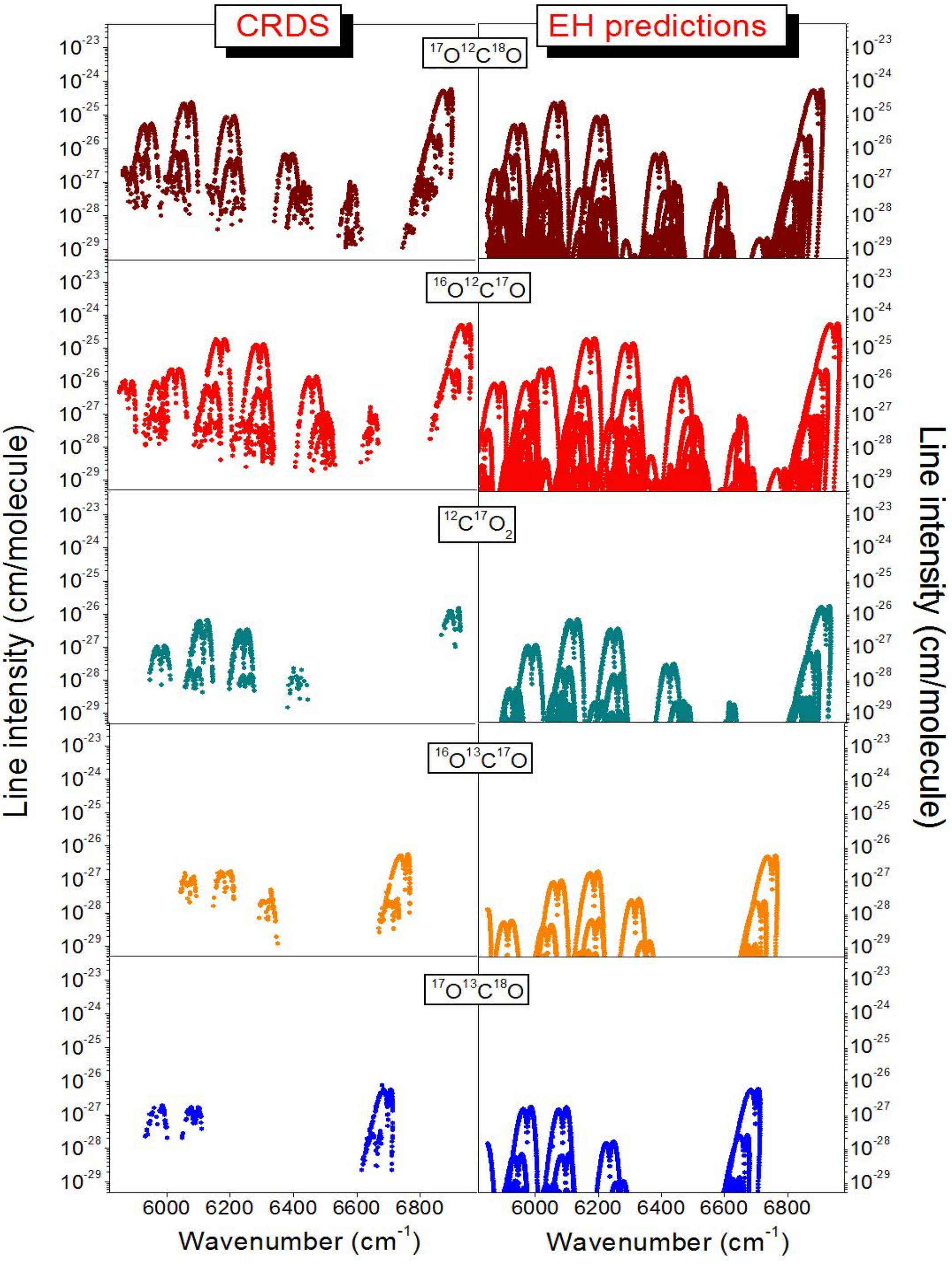

**Fig. 3.** Overview comparison between the transitions of $^{17}O^{12}C^{18}O$, $^{16}O^{12}C^{17}O$, $^{12}C^{17}O_2$, $^{16}O^{13}C^{17}O$ and $^{17}O^{13}C^{18}O$ assigned in the present study between 5850 and 7000 cm$^{-1}$ (left panels) and the corresponding spectra predicted within the framework of the effective operator approach (right panels). Line intensities correspond to the relative abundance of the various isotopologues in our sample (see Table 1).

Fig. 3 shows an overview comparison between the observed and predicted spectra of the five considered isotopologues. The experimental line lists are provided as Supplementary

Material. They include the rovibrational assignments and the positions and intensities calculated after a new fit of the EH and EDM parameters (see Sections 5 and 6). Table 3 compares to the present results the numbers of bands and transitions previously reported in the literature.

Overall, 1759 $^{16}O^{12}C^{17}O$ transitions belonging to 24 bands were assigned. The overall list is showed on Fig. 4 where previous observations by Fourier Transform Spectroscopy (FTS) of highly $^{17}O$ enriched carbon dioxide [13] and by CW-CRDS of natural carbon dioxide [14] are highlighted. Eight of these bands were previously reported by both FTS and CW-CRDS and three additional bands were reported only from CW-CRDS spectra. Among the newly observed bands, note the two perpendicular bands: 11121-00001 and 11122-00001 at 6646.94 and 6505.30 cm$^{-1}$, respectively. The new detection of these bands in the 0.20 Torr spectra (see Fig. 1) while the $^{16}O^{12}C^{17}O$ relative abundance is only about 2 % is a nice illustration of the sensitivity of the analyzed spectra. Fig. 4 shows that a large amount of the new lines are high rotational transitions of the previously observed bands.

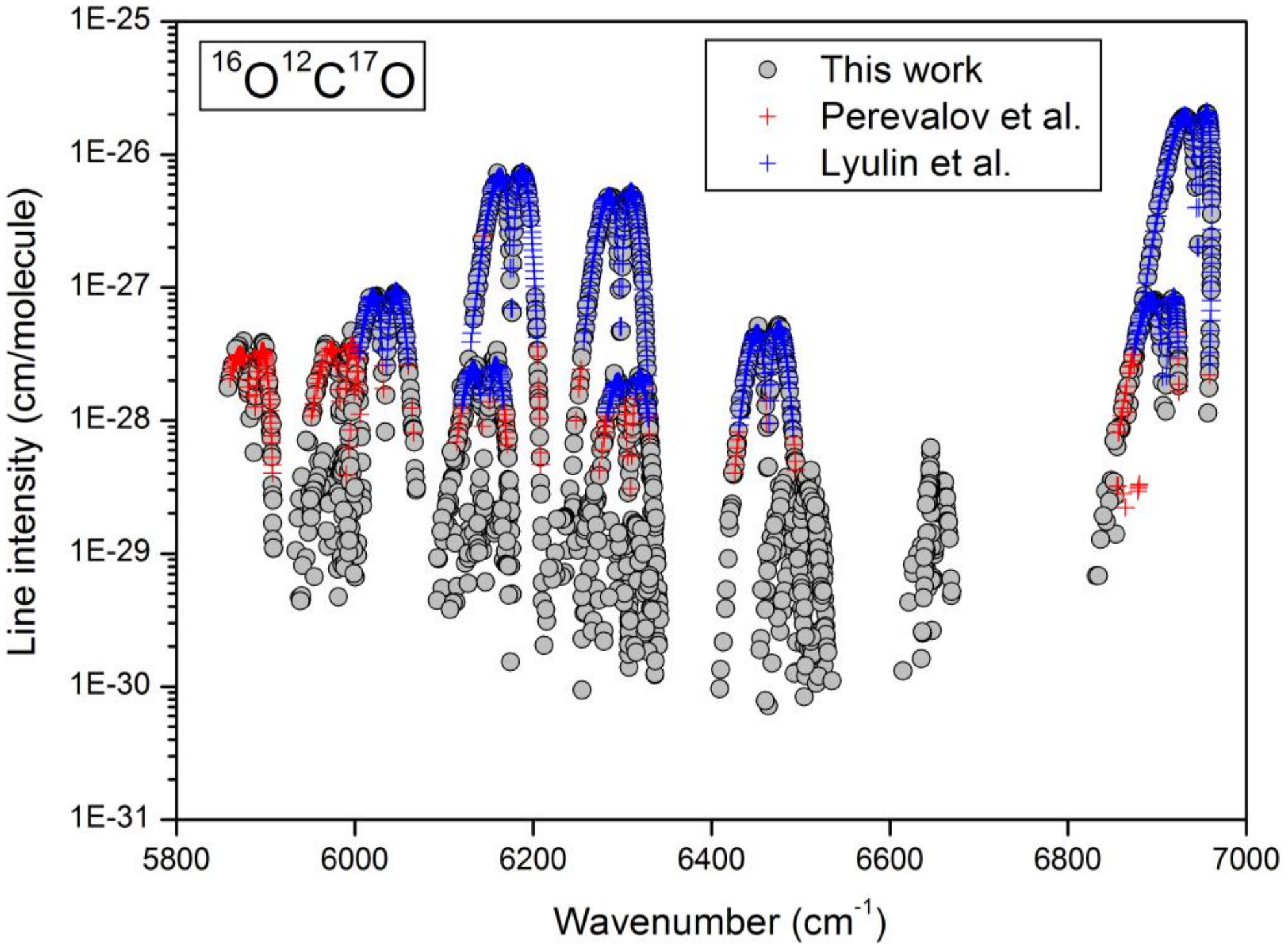

**Fig. 4.** Overview of the $^{16}O^{12}C^{17}O$ observations between 5850 and 7000 cm$^{-1}$. The previous observations by FTS and CW-CRDS [13,14] are highlighted (crosses) and superimposed to the present CW-CRDS measurements. Line intensities are given for $^{16}O^{12}C^{17}O$ in natural relative abundance (7.34×10$^{-4}$).

1786 transitions belonging to 24 bands were assigned to $^{17}O^{12}C^{18}O$ (Fig.5). Only five of them were previously reported from $^{17}O$ enriched carbon dioxide spectra: *(i)* the 3001*i*-00001 (*i*= 1-3) bands of the tetrad were studied using OPO–Femto-FT-CEAS [15,16], *(ii)* the 3001*i*-00001 (*i*= 2-4) bands of the same tetrad and the 00031-00001 band were measured by FTS [13].

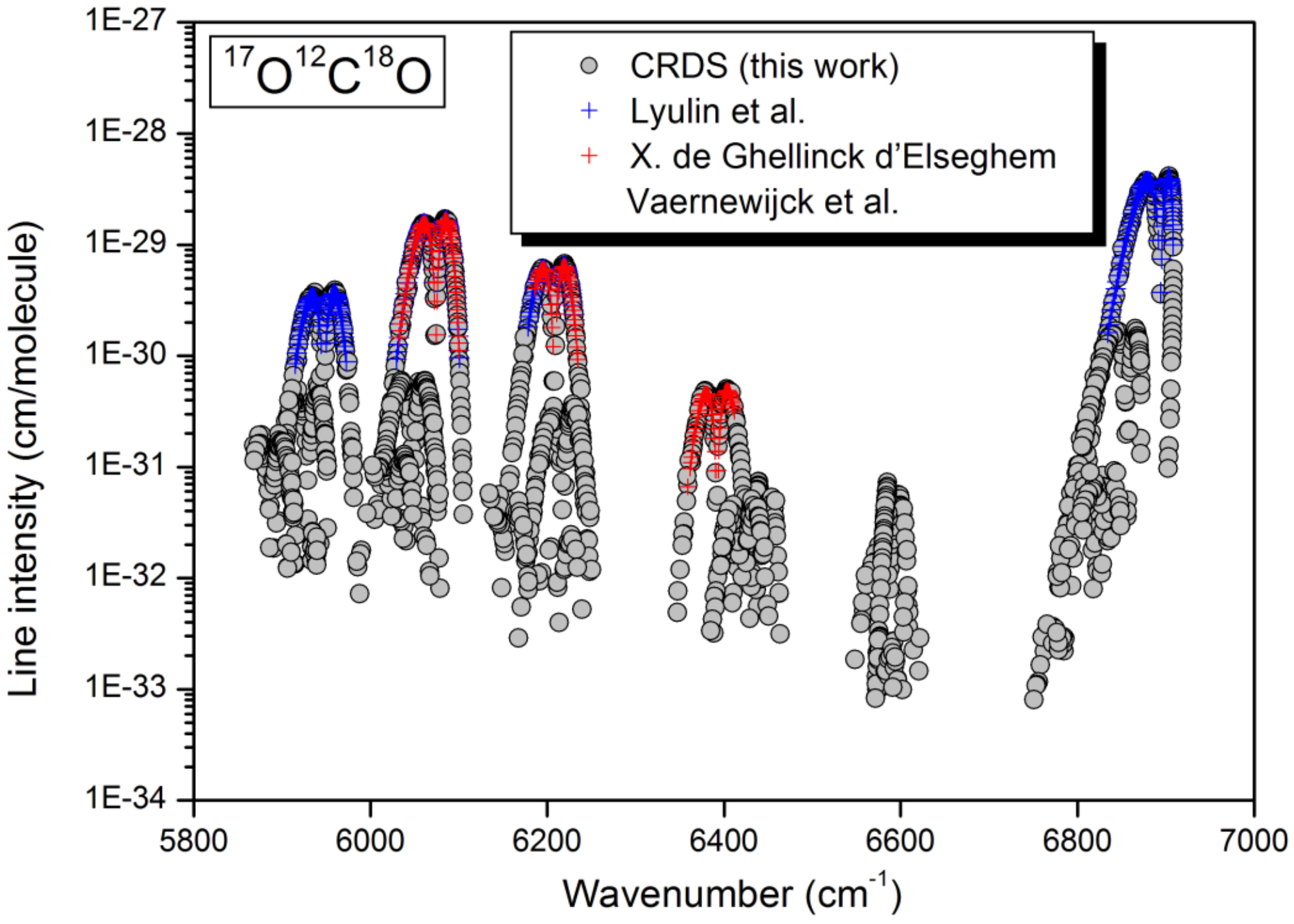

**Fig.5.** Overview of the $^{17}O^{12}C^{18}O$ observations between 5850 and 7000 cm$^{-1}$. The previous observations by FTS [13] (blue crosses) and OPO–Femto-FT-CEAS [15,16] (red crosses) are superimposed to the present CW-CRDS measurements. Line intensities are given for $^{17}O^{12}C^{18}O$ in natural relative abundance ($1.47\times10^{-6}$).

The $^{12}C^{17}O_2$, $^{16}O^{13}C^{17}O$ and $^{17}O^{13}C^{18}O$ isotopologues have a very small relative abundance in our sample (less than $5\times10^{-4}$). Only five $^{16}O^{13}C^{17}O$ bands (335 lines) and four $^{17}O^{13}C^{18}O$ bands (273 lines) were assigned. All assigned bands of $^{16}O^{13}C^{17}O$ were reported before using the same CW-CRDS spectrometer but with natural and $^{13}C$ enriched samples [10,17,18,19] (Fig. 6). In the case of $^{17}O^{13}C^{18}O$, three of the four assigned bands were reported by CW-CRDS in Ref. [17] and the 01131-01101 hot band is assigned for the first time (Fig. 7).

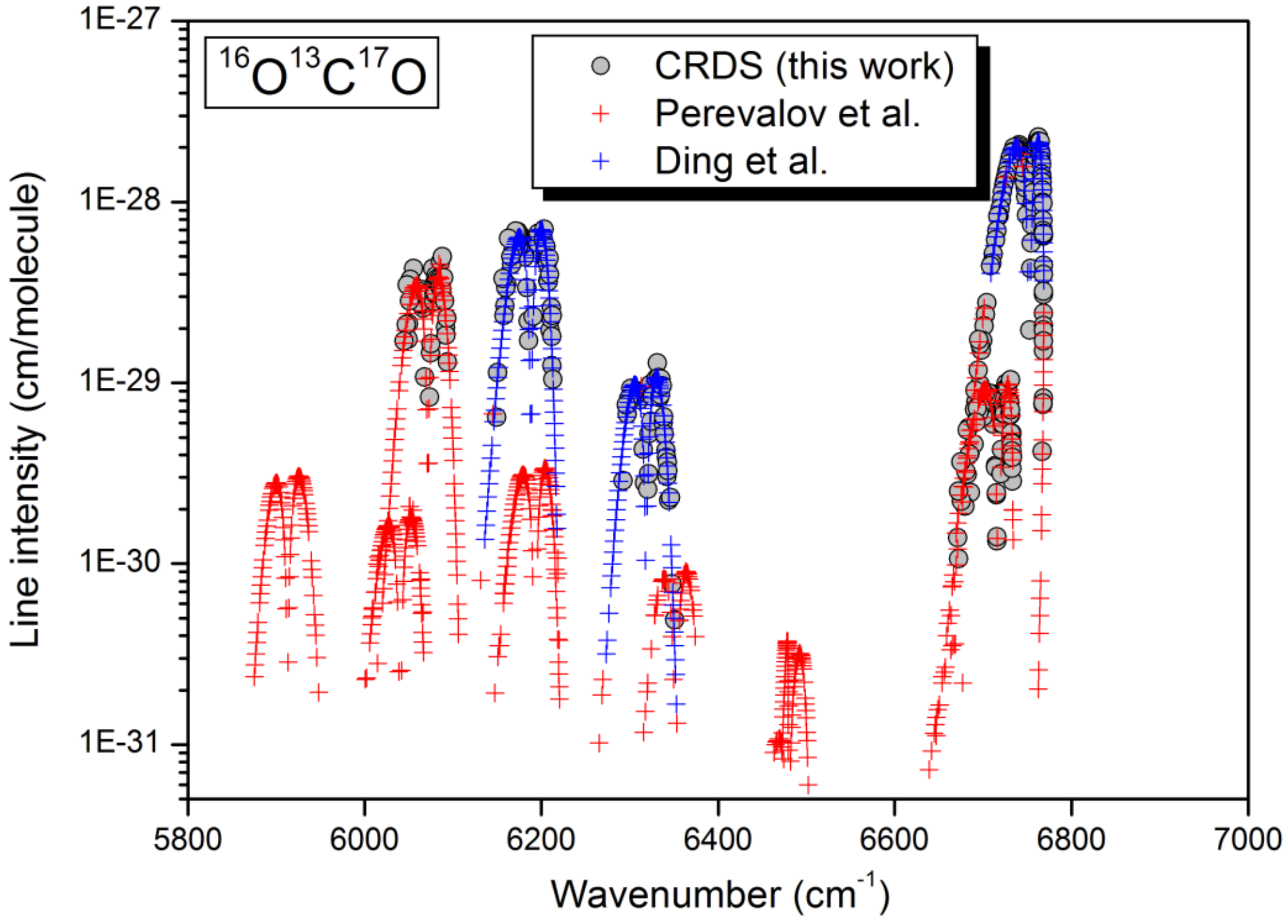

**Fig.6.** Overview of the $^{16}O^{13}C^{17}O$ observations between 5850 and 7000 cm$^{-1}$. The previous observations by FTS and CW-CRDS [10,17,18,19] are highlighted (crosses) and superimposed to the present CW-CRDS measurements. Line intensities are given for $^{16}O^{13}C^{17}O$ in natural relative abundance ($8.25\times10^{-6}$).

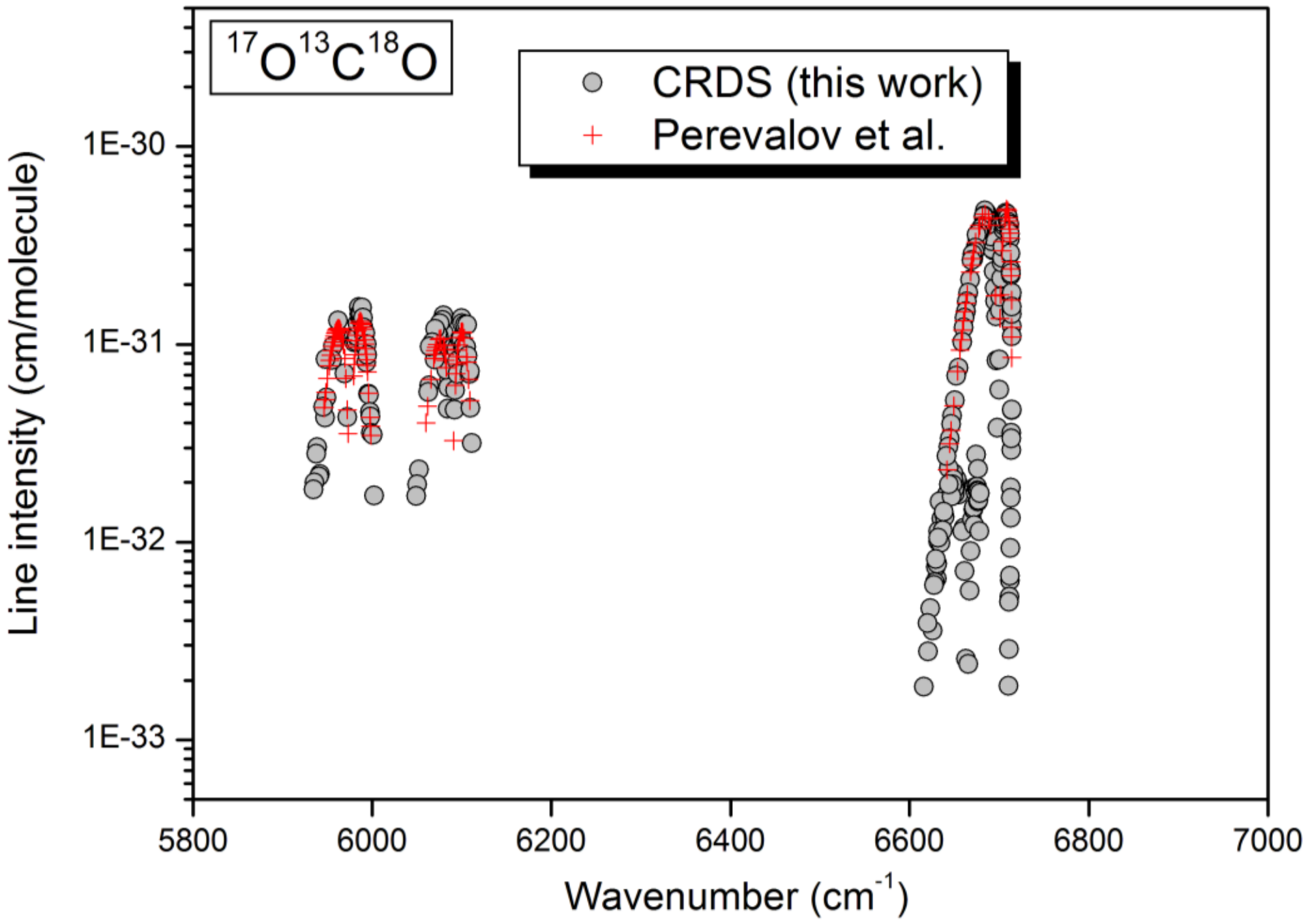

**Fig. 7.** Overview of the $^{17}O^{13}C^{18}O$ observations between 5850 and 7000 cm$^{-1}$. The previous observations by CW-CRDS [17] are highlighted (crosses) and superimposed to the present

CW-CRDS measurements. Line intensities are given for $^{17}O^{13}C^{18}O$ in natural relative abundance ($1.68\times10^{-8}$ [20]).

Before this work, seven bands of $^{12}C^{17}O_2$ were known in the 5850-7000 cm$^{-1}$ spectral region from FTS [13] and OPO–Femto-FT-CEAS spectra of $^{17}O$ enriched samples [15,16]. They are presented in Fig. 8. In addition, two hot bands of this isotopologue 3111*i*-01101 (*i*= 2, 3) were reported in Refs.[15,16] but our calculations show that their assignments are wrong (the reported band centers are shifted by about 9 and 20 cm$^{-1}$ from their correct values). As a result of a relative concentration on the order of $4\times10^{-4}$, only the four strongest of the previously known $^{12}C^{17}O_2$ bands, (30013-00001, 30012-00001, 30011-00001 and 00031-00001) were assigned in our spectra. We take advantage of the present work to complete our set of $^{12}C^{17}O_2$ line positions with recent assignments obtained in the same region from the analysis of our CW-CRDS spectrum of $^{17}O$ enriched water where $^{12}C^{17}O_2$ was present as an impurity [21]. Five $^{12}C^{17}O_2$ bands including the newly observed 31112-01101 band were assigned in the water spectrum. These data are included in Fig. 8 and used in the forthcoming band by band analysis.

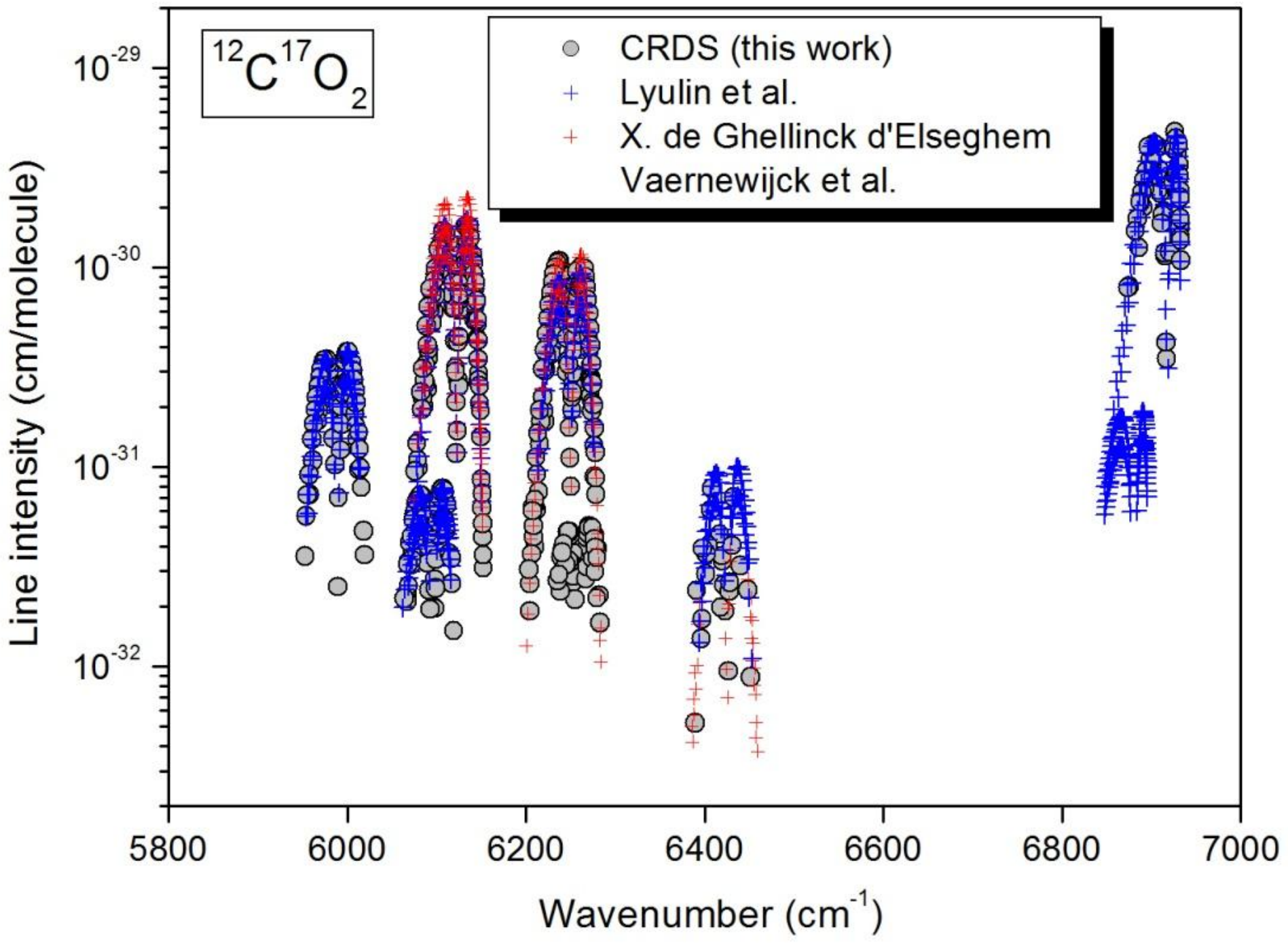

**Fig. 8.** Overview of the $^{12}C^{17}O_2$ observations between 5850 and 7000 cm$^{-1}$. The previous observations by FTS [13] (blue crosses) and OPO–Femto-FT-CEAS [15,16] (red crosses) are superimposed to the present CW-CRDS measurements. Line intensities are given for $^{12}C^{17}O_2$ in natural relative abundance ($1.37\times10^{-7}$). Calculated intensity values were used for the CRDS data obtained from $^{17}O$ enriched water spectra [21] because the abundance of $^{12}C^{17}O_2$ present as an impurity in water was unknown (see Text).

## Band-by-band analysis

As far as a vibrational state can be considered as isolated, the vibration-rotational energy levels can be expressed as:

$$F_v(J) = G_v + B_v J(J+1) - D_v J^2 (J+1)^2 + H_v J^3 (J+1)^3, \qquad (1)$$

where $J$ is the total angular momentum quantum number, $G_v$ is the vibrational term value, $B_v$ is the rotational constant, $D_v$ and $H_v$ are centrifugal distortion constants.

The spectroscopic parameters of the upper states of the different isotopologues were fitted to the measured line positions. The retrieved constants resulting from the band by band analysis procedure are listed in Tables 4-8 for $^{16}O^{12}C^{17}O$, $^{17}O^{12}C^{18}O$, $^{16}O^{13}C^{17}O$, $^{17}O^{13}C^{18}O$ and $^{12}C^{17}O_2$, respectively. The bands are ordered according to the upper vibrational term value. As the *e* and *f* sub bands may be perturbed in a different way, different sets of spectroscopic parameters were fitted for the *e* and *f* levels. The lower state rotational constants were constrained to the literature values [22,23,24]. (In the case of the 03331-03301 band of $^{17}O^{12}C^{18}O$, only a few lines were measured which prevented to fit the upper state constants. We provide in Table 5 the lower state constants predicted using the global effective Hamiltonian).

All *rms* values of the ($\nu_{obs}$-$\nu_{fit}$) deviations included in Tables 4-8 are within $10^{-3}$ cm$^{-1}$, which is very satisfactory considering the weakness of the considered transitions and our claimed accuracy of $1\times10^{-3}$ cm$^{-1}$ on the line positions. The list of measured line positions with assignments and ($\nu_{obs}$-$\nu_{fit}$) values is provided as Supplementary Material. This archive file includes for each band, the fitted spectroscopic parameters of the upper state with corresponding errors (in % and in cm$^{-1}$), lower state parameters and *rms* values of the fit together with the observed, calculated and EH values for the line positions.

Interpolyad anharmonic resonance perturbations were found to affect one band of the $^{16}O^{12}C^{17}O$ and $^{17}O^{12}C^{18}O$ asymmetric isotopologues (Table 9). The interpolyad anharmonic resonance interaction between the 30013 and 50006 vibrational states of $^{17}O^{12}C^{18}O$ was already discussed in Ref. [13].

**Table 9.** Interpolyad anharmonic resonance interactions affecting $^{16}O^{12}C^{17}O$ and $^{17}O^{12}C^{18}O$ bands between 5800 and 7000 cm$^{-1}$.

| Isotopologue | Band affected | Center (cm$^{-1}$) | Perturber | $J_{cross}$ [a] |
|---|---|---|---|---|
| $^{16}O^{12}C^{17}O$ | 11121e-00001e | 6646.9386 | 31114e | 39 |
| | 11121f-00001e | 6646.9396 | 31114f | 36 |
| $^{17}O^{12}C^{18}O$ | 30013e-00001e | 6073.7612 | 50006e | 33 |

Note
[a] Value of the angular momentum quantum number at which the energy level crossing takes place.

**Table 4.** Spectroscopic constants (in $cm^{-1}$) of the $^{16}O^{12}C^{17}O$ bands analyzed in the $^{18}O$ enriched spectrum of carbon dioxide recorded by CW-CRDS between 5851 and 6990 $cm^{-1}$.

| ***State*** ***($V_1V_2\ell_2V_3\, r\, \varepsilon$)*** | $G_v$ | $B_v$ | $D_v\times10^7$ | $H_v\times10^{12}$ | ***Ref.*** | | | | | | | |
|---|---|---|---|---|---|---|---|---|---|---|---|---|
| 00001e | 0. | 0.37861462 | 1.26428 | 0 | [22] | | | | | | | |
| 01101e | 664.72914 | 0.37902973 | 1.26216 | 0 | [22] | | | | | | | |
| 01101f | 664.72914 | 0.37961283 | 1.27387 | 0 | [22] | | | | | | | |
| 10002e | 1272.28663 | 0.37870010 | 1.46815 | 0 | [22] | | | | | | | |
| 02201e | 1329.843 | 0.380030 | 1.31 | 0 | [22] | | | | | | | |
| 02201f | 1329.843 | 0.380030 | 1.30 | 0 | [22] | | | | | | | |
| 10001e | 1376.02747 | 0.37877938 | 1.08312 | 0 | [22] | | | | | | | |
| ***State*** ***($V_1V_2\ell_2V_3\, r\, \varepsilon$)*** | $G_v$ | $B_v$ | $D_v\times10^7$ | $H_v\times10^{12}$ | **Fitted bands** | $P'$-$P''$ | $\Delta G_v$ [a] | observed lines [b] | *n/N* [c] | *RMS* [d] | *Note* [e] | Previous work [f] |
| ***P= 8*** | | | | | | | | | | | | |
| 10022e | 5885.32060(12) | 0.37292536(30) | 1.4789(11) | 0 | 10022e-00001e | 8-0 | 5885.32060(12) | P30/R52 | 61/71 | 0.51 | | P |
| 10021e | 5986.128713(90) | 0.37271694(24) | 1.0677(11) | 0 | 10021e-00001e | 8-0 | 5986.128713(90) | P50/R47 | 69/69 | 0.40 | | P |
| ***P= 9*** | | | | | | | | | | | | |
| 30014e | 6033.47614(10) | 0.37704212(33) | 1.9517(25) | 0.961(50) | 30014e-00001e | 9-0 | 6033.47614(10) | P61/R54 | 99/101 | 0.44 | | P,L |
| 30013e | 6175.951623(71) | 0.37498388(19) | 1.5501(10) | 1.109(14) | 30013e-00001e | 9-0 | 6175.951623(71) | P59/R70 | 105/106 | 0.36 | | P,L |
| 30012e | 6298.112530(69) | 0.37531443(16) | 0.91110(82) | 0.514(11) | 30012e-00001e | 9-0 | 6298.112530(69) | P68/R74 | 120/127 | 0.37 | | P,L |
| 30011e | 6463.480404(92) | 0.37675465(28) | 0.6948(19) | 0.636(33) | 30011e-00001e | 9-0 | 6463.480404(92) | P63/R64 | 107/115 | 0.44 | | P,L |
| 11122e | 6505.30254(11) | 0.37328131(32) | 1.4027(16) | 0 | 11122e-00001e | 9-0 | 6505.30254(11) | P45/R46 | 54/64 | 0.43 | | |
| 11122f | 6505.30280(16) | 0.37411531(38) | 1.4730(16) | 0 | 11122f-00001e | 9-0 | 6505.30280(16) | Q51 | 35/38 | 0.47 | | |
| 11121e | 6646.93855(19) | 0.37302948(57) | 1.1717(30) | 0 | 11121e-00001e | 9-0 | 6646.93855(19) | P34/R44 | 38/45 | 0.65 | 1 | |
| 11121f | 6646.93959(26) | 0.37383347(75) | 1.1315(37) | 0 | 11121f-00001e | 9-0 | 6646.93959(26) | Q48 | 35/39 | 0.74 | 2 | |
| 00031e | 6945.597515(88) | 0.36967212(14) | 1.25529(37) | 0 | 00031e-00001e | 9-0 | 6945.597515(88) | P68/R62 | 100/102 | 0.48 | | P,L |
| 11121e | 6646.93938(58) | 0.3730251(26) | 1.149(23) | 0 | 11121e-01101e | 9-1 | 5982.21024(58) | P15/R32 | 10/11 | 0.66 | | |
| 11121f | 6646.93871(64) | 0.3738402(43) | 1.196(44) | 0 | 11121f-01101e | 9-1 | 5982.20957(64) | P30/R28 | 9/10 | | | |
| ***P= 10*** | | | | | | | | | | | | |
| 31114e | 6641.33291(24) | 0.37673398(73) | 1.6139(43) | 0 | 31114e-01101e | 10-1 | 5976.60377(24) | P45/R39 | 31/35 | 0.64 | | |
| 31114f | 6641.33381(26) | 0.37831259(73) | 1.7912(37) | 0 | 31114f-01101f | 10-1 | 5976.60467(26) | P46/R37 | 30/34 | 0.65 | | |
| 31113e | 6810.89444(15) | 0.37534533(55) | 1.2067(47) | -2.03(10) | 31113e-01101e | 10-1 | 6146.16530(15) | P56/R55 | 69/76 | 0.51 | | P,L |
| 31113f | 6810.89385(16) | 0.37665538(39) | 1.3653(19) | 0 | 31113f-01101f | 10-1 | 6146.16471(16) | P47/R47 | 61/69 | 0.56 | | P,L |
| 31112e | 6972.728679(93) | 0.37546064(18) | 1.08669(59) | 0 | 31112e-01101e | 10-1 | 6307.999540(93) | P56/R59 | 82 /87 | 0.50 | | P,L |
| | | | | | 31112e-01101f | 10-1 | | Q16 | 5/11 | | | |
| 31112f | 6972.728468(71) | 0.37679681(15) | 1.05813(51) | 0 | 31112f-01101f | 10-1 | 6307.999330(71) | P58/R56 | 88/90 | 0.41 | | P,L |
| | | | | | 31112f-01101e | 10-1 | | Q13 | 7/8 | | | |
| 31111e | 7160.43924(15) | 0.37634425(40) | 0.9263(18) | 0 | 31111e-01101e | 10-1 | 6495.71010(15) | P47/R49 | 47/55 | 0.52 | | |
| 31111f | 7160.43994(18) | 0.37788965(51) | 0.8906(26) | 0 | 31111f-01101f | 10-1 | 6495.71080(18) | P41/R45 | 48/54 | 0.57 | | |
| 01131f | 7573.09230(21) | 0.37070963(32) | 1.2673(10) | 0 | 01131f-01101f | 10-1 | 6908.36316(21) | P60/R39 | 45/74 | 0.57 | 3 | P,L |
| 01131e | 7573.09249(22) | 0.37017054(54) | 1.2406(27) | 0 | 01131e-01101e | 10-1 | 6908.36335(22) | P47/R38 | 35/63 | 0.52 | | P,L |
| ***P= 11*** | | | | | | | | | | | | |
| 40014e | 7397.75813(27) | 0.37542785(88) | 1.8358(67) | 0 | 40014e-10002e | 11-2 | 6125.47150(27) | P37/R31 | 21/22 | 0.44 | | |
| 32213f | 7450.9696(14) | 0.3769769(84) | 1.34(10) | 0 | 32213f-02201f | 11-2 | 6121.1266(14) | P11/R26 | 8/8 | 0.95 | | |
| 40013e | 7525.82030(13) | 0.37434176(40) | 0.9809(22) | 0 | 40013e-10001e | 11-2 | 6149.79283(13) | P45/R44 | 26/32 | 0.52 | | |
| | | | | | 40013e-10002e | 11-2 | 6253.53367(13) | P44/R33 | 34/38 | | | |
| 32212e | 7641.74192(43) | 0.3769195(16) | 1.467(11) | 0 | 32212e-02201e | 11-2 | 6311.89892(43) | P37/R38 | 11/12 | 0.74 | | |
| 32212f | 7641.74270(46) | 0.3769104(16) | 1.1097(99) | 0 | 32212f-02201f | 11-2 | 6311.89970(46) | P37/R39 | 16/17 | | | |
| 40012e | 7676.42510(22) | 0.37590824(70) | 0.6817(38) | 0 | 40012e-10001e | 11-2 | 6300.39763(22) | P40/R45 | 34/42 | 0.69 | | |
| 32211e | 7851.0325(34) | 0.377802(35) | 1.85(67) | 0 | 32211e-02201e | 11-2 | 6521.1895(34) | P10/R19 | 7/7 | 0.93 | | |
| 32211f | 7851.0370(14) | 0.377499(24) | 1.22(82) | 0 | 32211f-02201f | 11-2 | 6521.1940(14) | P19/R33 | 6/6 | | | |
| 40011e | 7872.4203(18) | 0.3773072(60) | 0.415(42) | 0 | 40011e-10001e | 11-2 | 6496.3928(18) | R33 | 8/9 | 0.73 | | |
| 02231e | 8201.031377 [g] | | | | 02231e-02201e | 11-2 | 6871.188377 [g] | P36 | /6 | | 4 | P |
| 02231f | 8201.031332 [g] | | | | 02231f-02201f | 11-2 | 6871.188332 [g] | P36 | /6 | | | P |

*Notes*

The confidence interval (1 SD) between parentheses, is given the in the unit of the last quoted digit.

[a] $\Delta G_v = G_v^{'} - G_v^{''}$.

[b] Observed branches with the corresponding maximum value of the total angular momentum quantum number.

[c] *N* is the number of the observed lines for a given band and *n* is the number of these lines included in the fit.

[d] *RMS* of residuals of the spectroscopic parameters fit is given in $10^{-3}$ $cm^{-1}$.

[e] 1 - Interpolyad anharmonic interactions with 31114e (energy levels crossing at *J*= 39).

2 - Interpolyad anharmonic interactions with 31114f (energy levels crossing at *J*= 36).

3 - The e and f components are blended up to *J*=6 in the *P*-branch and up to *J*=29 in the *R*-branch. Unresolved lines were excluded from the fit.

4 - All the *e* and *f* components are not resolved.

[f] Previous reports: L: Ref. [13]; P: Ref. [14].

[g] A few lines observed, the given $G_v$, and $\Delta G_v$ values are EH values.

**Table 5.** Spectroscopic constants (in cm$^{-1}$) of the $^{17}O^{12}C^{18}O$ bands analyzed in the $^{18}O$ enriched spectrum of carbon dioxide recorded by CW-CRDS between 5851 and 6990 cm$^{-1}$.

| **State** $(V_1V_2\ell_2V_3\,r\,\varepsilon)$ | $G_v$ | $B_v$ | $D_v\times10^7$ | $H_v\times10^{12}$ | ***Ref.*** | | | | | | | |
|---|---|---|---|---|---|---|---|---|---|---|---|---|
| 00001e | 0. | 0.356931872 | 1.11566 | 0 | [23] | | | | | | | |
| 01101e | 659.701654 | 0.357347188 | 1.13217 | 0 | [23] | | | | | | | |
| 01101f | 659.701654 | 0.357868213 | 1.13798 | 0 | [23] | | | | | | | |
| 10002e | 1244.594135 | 0.356737834 | 1.26346 | 0.192 | [23] | | | | | | | |
| 02201e | 1319.818 | 0.358276600 | 1.19470 | -0.245 | [23] | | | | | | | |
| 02201f | 1319.818 | 0.358276600 | 1.15598 | 0.049 | [23] | | | | | | | |
| 10001e | 1355.654216 | 0.357407042 | 0.97108 | 0.210 | [23] | | | | | | | |
| 03301e | 1980.2757 [a] | 0.35894276 | 1.19206 | 0 | This work [a] | | | | | | | |
| 03301f | 1980.2757 [a] | 0.35894255 | 1.19057 | 0 | This work [a] | | | | | | | |
| **State** $(V_1V_2\ell_2V_3\,r\,\varepsilon)$ | $G_v$ | $B_v$ | $D_v\times10^7$ | $H_v\times10^{12}$ | Fitted bands | $P'$-$P''$ | $\Delta G_v$ [b] | Observed lines [c] | *n/N* [d] | *RMS* [e] | Note [f] | Previous work [g] |
| ***P= 8*** | | | | | | | | | | | | |
| 10021e | 5930.28915(12) | 0.35171833(37) | 0.9449(20) | 0 | 10021e-00001e | 8-0 | 5930.28915(12) | P46/R43 | 64/64 | 0.51 | | |
| ***P= 9*** | | | | | | | | | | | | |
| 30014e | 5946.97861(10) | 0.35482043(30) | 1.6688(21) | 0.472(37) | 30014e-00001e | 9-0 | 5946.97861(10) | P65/R59 | 108/109 | 0.49 | | L |
| 30013e | 6073.761161(97) | 0.35331366(28) | 1.1968(17) | 1.009(26) | 30013e-00001e | 9-0 | 6073.761161(97) | P65/R67 | 114/123 | 0.49 | 1 | L,$X_1$,$X_2$ |
| 30012e | 6207.76048(10) | 0.35441221(25) | 0.8035(15) | 0.3345(23) | 30012e-00001e | 9-0 | 6207.76048(10) | P68/R66 | 104/106 | 0.43 | | L, $X_2$ |
| 30011e | 6391.96188(11) | 0.35566762(21) | 0.55532(78) | 0 | 30011e-00001e | 9-0 | 6391.96188(11) | P58/R52 | 94/96 | 0.56 | | $X_2$ |
| 11122e | 6440.84194(17) | 0.35174275(39) | 1.2280(16) | 0 | 11122e-00001e | 9-0 | 6440.84194(17) | P35/R53 | 47/53 | 0.63 | | |
| 11122f | 6440.84247(13) | 0.35244951(32) | 1.2846(14) | 0 | 11122f-00001e | 9-0 | 6440.84247(13) | Q50 | 35/35 | 0.38 | | |
| 11121e | 6586.13178(14) | 0.35191686(34) | 1.0237(15) | 0 | 11121e-00001e | 9-0 | 6586.13178(14) | P41/R50 | 55/56 | 0.55 | | |
| 11121f | 6586.13135(10) | 0.35269105(19) | 1.00843(65) | 0 | 11121f-00001e | 9-0 | 6586.13135(10) | Q56 | 39/39 | 0.30 | | |
| 00031e | 6893.712316(70) | 0.348514663 (79) | 1.10796(16) | 0 | 00031e-00001e | 9-0 | 6893.712316(70) | P81/R67 | 123/123 | 0.43 | | L |
| 11121e | 6586.13152(76) | 0.3519191(32) | 1.019(30) | 0 | 11121e-01101e | 9-1 | 5926.42987(76) | P25/R29 | 12/12 | 0.53 | | |
| 11121f | 6586.13151(27) | 0.3526754(18) | 0.984(25) | 0 | 11121f-01101f | 9-1 | 5926.42987(27) | P20/R29 | 9/9 | 0.68 | | |
| ***P= 10*** | | | | | | | | | | | | |
| 31114e | 6547.22988(23) | 0.35478555(79) | 1.4168(46) | 0 | 31114e-01101e | 10-1 | 5887.52823(23) | P25/R42 | 41/41 | 0.71 | | |
| 31114f | 6547.22893(22) | 0.3560364(12) | 1.679(17) | 0.303(58) | 31114f-01101f | 10-1 | 5887.52728(22) | P25/R44 | 51/54 | 0.67 | | |
| 31113e | 6708.65831(15) | 0.35384052(53) | 1.0881(40) | -1.074(82) | 31113e-01101e | 10-1 | 6048.95666(15) | P50/R59 | 76/76 | 0.62 | | |
| 31113f | 6708.65764(13) | 0.35497236(32) | 1.1716(16) | 0 | 31113f-01101f | 10-1 | 6048.95599(13) | P47/R50 | 71/75 | 0.57 | | |
| 31112e | 6878.33146(13) | 0.35436302(28) | 0.9723(10) | 0 | 31112e-01101e | 10-1 | 6218.62981(13) | P58/R52 | 76/76 | 0.63 | | |
| 31112f | 6878.33112(11) | 0.35562924(25) | 0.9456(10) | 0 | 31112f-01101f | 10-1 | 6218.62947(11) | P51/R52 | 64/67 | 0.47 | | |
| 31111e | 7081.59558(20) | 0.35526211(73) | 0.8564(47) | 0 | 31111e-01101e | 10-1 | 6421.89393(20) | P41/R32 | 38/44 | 0.58 | | |
| 31111f | 7081.59570(16) | 0.35675307(42) | 0.8359(22) | 0 | 31111f-01101f | 10-1 | 6421.89405(16) | P48/R42 | 42/49 | 0.51 | | |
| 12221e | 7237.54470(51) | 0.3529443(21) | 1.195(16) | 0 | 12221e-01101e | 10-1 | 6577.84305(51) | R23 | 32/33 | 0.64 | | |
| | | | | | 12221e-01101f | 10-1 | | Q35 | | | | |
| 12221f | 7237.54498(41) | 0.3529375(17) | 0.976(14) | 0 | 12221f-01101f | 10-1 | 6577.84333(41) | R34 | | | | |
| | | | | | 12221f-01101e | 10-1 | | Q30 | | | | |
| 20021e | 7265.4792(11) | 0.3523380(65) | 0.790(72) | 0 | 20021e-01101e | 10-1 | 6605.7775(11) | P18/R27 | 8/9 | 0.91 | | |
| | | | | | 20021e-01101f | 10-1 | | Q21 | | | | |
| 01131e | 7516.56633(17) | 0.34900390(27) | 1.12219 (75) | 0 | 01131e-01101e | 10-1 | 6856.86468(17) | P64/R46 | 72/81 | 0.73 | | |
| | | | | | 01131e-01101f | 10-1 | | Q7 | | | | |
| 01131f | 7516.56728(17) | 0.34948656(26) | 1.13196(67) | 0 | 01131f-01101f | 10-1 | 6856.86563(17) | P67/R37 | 62/74 | 0.69 | | |
| ***P=11*** | | | | | | | | | | | | |
| 40014e | 7271.58940(21) | 0.35326911(64) | 1.4634(32) | 0 | 40014e-10002e | 11-2 | 6026.99527(21) | P48/R34 | 43/48 | 0.73 | | |
| 40013e | 7399.61505(22) | 0.35351291(70) | 0.7757(39) | 0 | 40013e-10002e | 11-2 | 6155.02092(22) | P26/R40 | 42/43 | 0.64 | | |
| | | | | | 40013e-10001e | 11-2 | 6043.96083(22) | P14/R42 | | | | |
| 40012e | 7570.71131(40) | 0.3550143(17) | 0.574(12) | 0 | 40012e-10001e | 11-2 | 6215.05709(40) | P32/R38 | 18/19 | 0.92 | | |
| 10032e | 8078.76170(28) | 0.3485594(11) | 1.3017(84) | 0 | 10032e-10002e | 11-2 | 6834.16757(28) | P31/R34 | 24/26 | 0.45 | | |
| 02231e | 8139.703406(12) [b] | | | | 02231e-02201e | 11-2 | 6819.885406(12) | P39/R35 | /43 | | 2 | |
| | | | | | 02231e-02201f | 11-2 | | Q4 | /3 | | | |
| 02231f | 8139.703366(12) [b] | | | | 02231f-02201f | 11-2 | 6819.885366(12) | P39/R35 | /43 | | | |
| | | | | | 02231f-02201e | 11-2 | | Q4 | /3 | | | |
| 10031e | 8181.20276(32) | 0.34886839(76) | 0.9427(30) | 0 | 10031e-10001e | 11-2 | 6825.54854(32) | P52/R25 | 24/24 | 0.73 | | |
| ***P=12*** | | | | | | | | | | | | |
| 03331e | 8763.307354 [a] | | | | 03331e-03301e | 12-3 | 6783.031668 [a] | P33 | /11 | | 2 | |
| 03331f | 8763.307362 [a] | | | | 03331f-03301f | 12-3 | 6783.031656 [a] | P33 | /11 | | | |

*Notes*
The confidence interval (1 SD) between parentheses, is given the in the unit of the last quoted digit.
[a] The given parameter values were obtained from the EH predictions.
[b] $\Delta G_v = G_v^{'} - G_v^{''}$.
[c] Observed branches with the corresponding maximum value of the total angular momentum quantum number.
[d] *N* is the number of the observed lines for a given band and *n* is the number of these lines involved in the fit.
[e] Root Mean Squares of residuals of the spectroscopic parameters fit is given in $10^{-3}$ $cm^{-1}$.
[f] 1 - Interpolyad anharmonic interaction with 50006e (energy levels crossing at *J*= 33).
2 - All the e and f components are not resolved.
[g] Previous reports: L: Ref.[13]; $X_1$: Ref.[15]; $X_2$: Ref.[16].

**Table 6.** Spectroscopic constants (in $cm^{-1}$) of the $^{16}O^{13}C^{17}O$ bands analyzed in the $^{18}O$ enriched spectrum of carbon dioxide recorded by CW-CRDS between 5851 and 6990 $cm^{-1}$.

| **State ($V_1V_2\ell_2V_3\, r\, \varepsilon$)** | $G_v$ | $B_v$ | $D_v \times 10^7$ | ***Ref.*** | | | | | | |
|---|---|---|---|---|---|---|---|---|---|---|
| 00001e | 0. | 0.37861700 | 1.24220 | [22] | | | | | | |
| 01101e | 645.744 | 0.378961 | 1.22 | [22] | | | | | | |
| 01101f | 645.744 | 0.379600 | 1.23 | [22] | | | | | | |
| **State ($V_1V_2\ell_2V_3\, r\, \varepsilon$)** | $G_v$ | $B_v$ | $D_v \times 10^7$ | Fitted bands | $P'$-$P''$ | $\Delta G_v$ [a] | Observed lines [b] | *n/N* [c] | *RMS* [d] | Previous work [e] |
| ***P= 9*** | | | | | | | | | | |
| 30013e | 6071.57304(21) | 0.3757853(10) | 1.6641(93) | 30013e-00001e | 9-0 | 6071.57304(21) | P32/R33 | 43/43 | 0.62 | P2,P3 |
| 30012e | 6188.05195(17) | 0.37459042(50) | 0.9681(28) | 30012e-00001e | 9-0 | 6188.05195(17) | P42/R42 | 52/52 | 0.56 | P2,D |
| 30011e | 6318.43282(17) | 0.37592903(43) | 0.7008(19) | 30011e-00001e | 9-0 | 6318.43282(17) | P31/R51 | 46/46 | 0.61 | P2,D |
| 00031e | 6752.412227(86) | 0.37000461(15) | 1.23279(51) | 00031e-00001e | 9-0 | 6752.412227(86) | P63/R56 | 95/95 | 0.45 | P1,P2,D,P3 |
| ***P= 10*** | | | | | | | | | | |
| 01131e | 7363.26690(28) | 0.3704379(12) | 1.2195(82) | 01131e-01101e | 10-1 | 6717.52290(28) | P41/R35 | 44/52 | 0.99 | P1,P2 |
| 01131f | 7363.26726(19) | 0.37102467(66) | 1.2173(46) | 01131f-01101f | 10-1 | 6717.52326(19) | P41/R35 | 47/47 | 0.59 | P1,P2 |

*Notes*

The confidence interval (1 SD) between parentheses, is given the in the unit of the last quoted digit. The $H_v$ constants were constrained to 0 both for the lower and upper states.

[a] $\Delta G_v = G_v^{'} - G_v^{''}$.

[b] Observed branches with the corresponding maximum value of the total angular momentum quantum number.

[c] *N* is the number of the observed lines for a given band and *n* is the number of these lines included in the fit.

[d] *RMS* of residuals of the spectroscopic parameters fit is given in $10^{-3}$ $cm^{-1}$.

[e] Previous reports: P1: Ref.[10]; P2: Ref.[17]; D: Ref.[18]; P3: Ref.[19].

**Table 7.** Spectroscopic constants (in cm$^{-1}$) of the $^{17}O^{13}C^{18}O$ bands analyzed in the $^{18}O$ enriched spectrum of carbon dioxide recorded by CW-CRDS between 5851 and 6990 cm$^{-1}$.

| **State** ($V_1V_2\ell_2V_3r\varepsilon$) | $G_v$ | $B_v$ | $D_v\times10^7$ | $H_v\times10^{12}$ | ***Ref.*** | | | | | | | |
|---|---|---|---|---|---|---|---|---|---|---|---|---|
| 00001e | 0. | 0.35694456 | 1.1113 | -0.18 | [24] | | | | | | | |
| 01101e | 640.56384 | 0.35731753 | 1.1259 | -0.18 | [24] | | | | | | | |
| 01101f | 640.56384 | 0.35785460 | 1.1398 | -0.18 | [24] | | | | | | | |
| **State** ($V_1V_2\ell_2V_3r\varepsilon$) | $G_v$ | $B_v$ | $D_v\times10^7$ | $H_v\times10^{12}$ | Fitted bands | $P'$-$P''$ | $\Delta G_v$ [a] | Observed lines [b] | *n/N* [c] | *RMS* [d] | Note [e] | Previous work [f] |
| ***P= 9*** | | | | | | | | | | | | |
| 30013e | 5975.17564(30) | 0.3537989(12) | 1.514(13) | 4.53(37) | 30013e-00001e | 9-0 | 5975.17564(30) | P47/R48 | 47/47 | 0.66 | | P |
| 30012e | 6088.67228(24) | 0.35372482(76) | 0.8058(40) | 0 | 30012e-00001e | 9-0 | 6088.67228(24) | P46/R38 | 42/44 | 0.82 | | P |
| 00031e | 6698.86089(11) | 0.34883973(18) | 1.11138(50) | 0 | 00031e-00001e | 9-0 | 6698.86089(11) | P64/R63 | 102/103 | 0.64 | | P |
| ***P= 10*** | | | | | | | | | | | | |
| 01131f | 7304.92223(60) | 0.3497812(15) | 1.1285(73) | 0 | 01131f-01101f | 10-1 | 6664.35839(60) | P45/R31 | 16/40 | 0.74 | 1 | |
| 01131e | 7304.92263(58) | 0.3492843(21) | 1.102(15) | 0 | 01131e-01101e | 10-1 | 6664.35879(58) | P37/R31 | 16/39 | 0.74 | | |

*Notes*

The confidence interval (1 SD) between parentheses, is given the in the unit of the last quoted digit.

[a] $\Delta G_v = G_v' - G_v''$.

[b] Observed branches with the corresponding maximum value of the total angular momentum quantum number.

[c] *N* is the number of the observed lines for a given band and *n* is the number of these lines included in the fit.

[d] *RMS* of residuals of the spectroscopic parameters fit is given in 10$^{-3}$ cm$^{-1}$.

[e] 1 - The e and f components are blended up to $J$ = 7 in the *P* branch and up to $J$ = 31 in the *R*-branch. Unresolved lines were excluded from the fit.

[f] Previous reports: P: Ref.[17].

**Table 8.** Spectroscopic constants (in $cm^{-1}$) of the $^{12}C^{17}O_2$ bands analyzed in the $^{18}O$ enriched spectrum of carbon dioxide recorded by CW-CRDS between 5851 and 6990 $cm^{-1}$.

| **State** **($V_1V_2\,\ell_2V_3\,r\,\varepsilon$)** | $G_v$ | $B_v$ | $D_v\times10^7$ | $H_v\times10^{12}$ | ***Ref.*** | | | | | | |
|---|---|---|---|---|---|---|---|---|---|---|---|
| 00001e | 0. | 0.36719462(82) | 1.18053(16) | 0 | [23] | | | | | | |
| 01101e | 662.066499(28) | 0.367612037(80) | 1.19784(34) | 0 | [23] | | | | | | |
| 01101f | 662.066499(28) | 0.368161373(80) | 1.20396(34) | 0 | [23] | | | | | | |
| ***State*** ***($V_1V_2\,\ell_2V_3\,r\,\varepsilon$)*** | $G_v$ | $B_v$ | $D_v\times10^7$ | $H_v\times10^{12}$ | Fitted bands | $P'$-$P''$ | $\Delta G_v$ [a] | Observed lines [b] | *n/N* [c] | *RMS* [d] | Previous work[e] |
| ***P= 9*** | | | | | | | | | | | |
| 30014e | 5988.91375(20) | 0.36534838(52) | 1.7689(26) | 0 | 30014e-00001e | 9-0 | 5988.91375(20) | P44/R47 | 58/63 | 0.76 | L |
| 30013e | 6122.70465(10) | 0.36352850(37) | 1.3565(30) | 1.039(66) | 30013e-00001e | 9-0 | 6122.70465(10) | P57/R57 | 141/147 | 0.56 | L,$X_1$, $X_2$ |
| 30012e | 6249.95103(91) | 0.36432177(21) | 0.82830(85) | 0 | 30012e-00001e | 9-0 | 6249.95103(91) | P53/R55 | 141/150 | 0.59 | L, $X_2$ |
| 30011e | 6425.20573(29) | 0.36566372(95) | 0.6119(55) | 0 | 30011e-00001e | 9-0 | 6425.20573(29) | P45/R38 | 24/24 | 0.82 | L, $X_2$ |
| 00031e | 6917.81966(15) | 0.35852830(55) | 1.1730(38) | 0 | 00031e-00001e | 9-0 | 6917.81966(15) | P41/R36 | 50/54 | 0.53 | L |
| ***P= 10*** | | | | | | | | | | | |
| 31113e | 6757.44051(27) | 0.3640028(11) | 1.1289(89) | 0 | 31113e-01101e | 10-1 | 6095.37401(27) | P33/R36 | 33/36 | 0.71 | L |
| 31113f | 6757.44029(32) | 0.3652101(13) | 1.2113(89) | 0 | 31113f-01101f | 10-1 | 6095.37379(32) | P37/R38 | 28/34 | 0.75 | L |
| 31112e | 6922.67610(46) | 0.3643590(20) | 1.069(15) | 0 | 31112e-01101e | 10-1 | 6260.60960(46) | P30/R36 | 20/22 | 0.94 | |
| 31112f | 6922.67703(66) | 0.3656552(39) | 1.016(45) | 0 | 31112f-01101f | 10-1 | 6260.61053(66) | P29/R27 | 20/21 | 0.88 | |

*Notes*

The confidence interval (1 SD) between parentheses, is given the in the unit of the last quoted digit.

The input data was extended by using some positions of $^{12}C^{17}O_2$ lines assigned in a CRDS spectrum of $^{17}O$ enriched water [21].

[a] $\Delta G_v = G_v' - G_v''$.

[b] Observed branches with the corresponding maximum value of the total angular momentum quantum number.

[c] *N* is the number of the observed lines for a given band and *n* is the number of these lines included in the fit.

[d] *RMS* of residuals of the spectroscopic parameters fit is given in $10^{-3}$ $cm^{-1}$.

[e] Previous reports: L: Ref.[13]; $X_1$: Ref.[15]; $X_2$: Ref.[16].

## 5. Global fitting of the line positions

Our measured line positions were gathered with all the position measurements available in the literature and used for the refinement of the EH parameters of three isotopologues: $^{17}O^{12}C^{18}O$, $^{16}O^{13}C^{17}O$ and $^{17}O^{13}C^{18}O$. For the $^{16}O^{12}C^{17}O$ and $^{12}C^{17}O_2$ species, the position values obtained in this work were already included as input data for the global fit of the respective EH parameters [25]. For the global modeling of the line positions, the EH and GIP computer codes were used [6,7,8]. When necessary, the input position values of some experimental sources were corrected using the correction factors determined with the help of the RITZ computer code [26], applied simultaneously to all collected sources of carbon dioxide line positions. The initial sets of the EH parameters were obtained by isotopic extrapolation of that of the principal isotopologue [27]. The weighted fits were performed taking into account of the claimed uncertainties of the line position determination in the respective references. In the cases when these uncertainties are not presented clearly, the weighting was based on the reported *rms* errors of the band by band fits of the spectroscopic constants. The fitted EH parameters of the three considered isotopologues and their standard errors are provided as Supplementary Material.

### $^{17}O^{12}C^{18}O$

The existing set of EH parameters of $^{17}O^{12}C^{18}O$ was published by Chedin in 1979 [28]. The effective Hamiltonian used in Ref. [29] was formulated only up to the fourth order of perturbation theory and some important Coriolis and anharmonic resonance interaction terms were not taken into account. The predictive ability of this set of parameters is on the order of 0.01 - 1 $cm^{-1}$ depending on the vibrational excitation.

The input file of the measured line positions used in our fit (Table 10) consists of the data published in Refs. [13,15,16,23,25,29,30,31,32,33] and data from this work. Overall, 9754 line positions were measured between 571 and 8197 $cm^{-1}$. The maximal value of the vibrational normal mode excitation is 4 for all modes. The highest value of the upper state rotational quantum number *J* is 98. In the first column, the sources of the input data are presented. A given reference may be attached to several sources according to the sets of spectra analyzed. The second column indicates the type of gas samples used for the recordings, the third column lists isotopologues found in the samples, and the fourth column specifies the experimental technique. Statistically significant frequency correction factors found from the Ritz analysis of the most abundant isotopologues are given in column 5. Columns 6 and 7 contain numbers of accepted and excluded transitions, respectively. Wavenumber range of each source is given in column 8. Minimal and maximal measurement uncertainties are given in column 9 in units of $10^{-3}$ $cm^{-1}$. Finally, *RMS* errors for each source are listed in column 10 in units of $10^{-3}$ $cm^{-1}$ ($RMS_{RITZ}$). The four last rows of the table contain number of input line positions, number of energy levels used to fit line positions, *RMS* of the residuals in units of $10^{-3}$ $cm^{-1}$, and the weighted dimensionless standard deviation $\chi$ of the fit. The data rejected by the RITZ analysis were excluded from the fits.

4139 energy levels are involved in the data file of 9754 measured line positions used to fit the EH parameters. They can be partitioned into $N_s$ independent sets or spectroscopic networks, each of them being connected internally by transitions of the data file (similarly to the MARVEL approach developed in Ref. [34]) . The largest set includes 3563 energy levels which are connected with 9456 transitions. The 576 remaining energy levels corresponding to 298 measured transitions are scattered on a large number of small spectroscopic networks which were not analysed.

During the fitting process, 63 parameters were varied. The resulting weighted dimensionless standard deviation of the fit is 1.40 and the *RMS* of the residuals is 0.00108 $cm^{-1}$. The ($\nu_{obs}$-$\nu_{EH}$) residuals corresponding to the line positions used as input data are plotted in the upper panel of Fig. 9, together with measurement uncertainties.

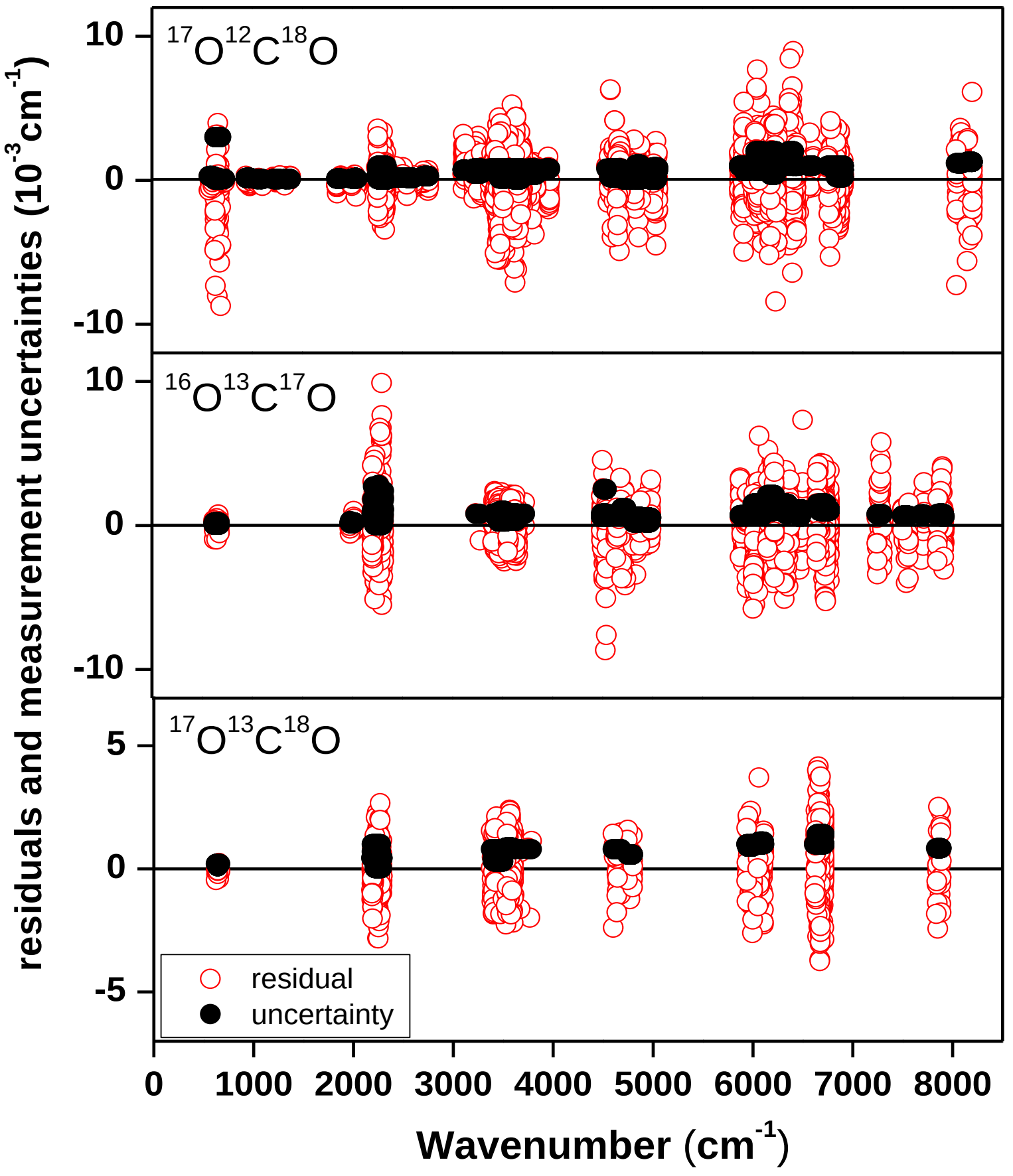

**Fig. 9.** Residuals (open circles) and measurement (full circles) uncertainties of the line positions used as input data for the fit of the effective Hamiltonian parameters of $^{17}O^{12}C^{18}O$, $^{16}O^{13}C^{17}O$ and $^{17}O^{13}C^{18}O$.

## $^{16}O^{13}C^{17}O$

The set of the EH parameters derived by Chedin [28] was refined in our previous paper [10] using published line positions from Refs. [10,18,19,35,36,37,38,39]. The refined set of parameters was used for the calculation of the line positions included in the current version of the CDSD-296 databank [40]. Since that time a number of new laboratory measurements have become available [13,17,25,33,41,42].

The new fit of the effective Hamiltonian parameters involved all published data [13,17,18,19,25,35,36,37,38,39,41,42] and data obtained in this work (see Table 11). Overall, 4258 line positions were reported between 617 and 7917 cm$^{-1}$. The number of involved energy levels is 2339. The highest value of the upper state rotational quantum number $J$ is 82. Nineteen lines of the 40014-00001 band had to be excluded from the EH fit because the 40014 upper vibrational state is strongly perturbed by an interpolyad anharmonic resonance interaction with the 60007 vibrational state [42]. The used polyad model of effective Hamiltonian does not take into account this type of resonance interactions.

During the fitting process, 43 parameters were varied. The resulting weighted dimensionless standard deviation of the fit is 1.36 and the *RMS* of the residuals is 0.00117 cm$^{-1}$. The fitted EH parameters and their standard errors are given as Supplementary Material. The ($\nu_{obs}$-$\nu_{EH}$) residuals together with measurement uncertainties are plotted in the middle panel of Fig. 9.

## $^{17}O^{13}C^{18}O$

A first set of EH parameters allowing for the calculation of the line positions with the accuracy on the order of 0.01-1 cm$^{-1}$ was published by Chedin and Teffo [43] for this isotopologue. The set of $^{17}O^{13}C^{18}O$ parameters was recently determined by isotopic extrapolation from the principal isotopologue [11]. In the present work, the EH parameters of $^{17}O^{13}C^{18}O$ are fitted for the first time.

The input file of the measured line positions consists of the data published in Refs. [13,17,25,39,42] and CRDS data from this work (see Table 12). Overall, 2008 line positions were measured between 623 and 7887 cm$^{-1}$. The maximal value of the vibrational normal mode excitation is 3 for all modes. The highest value of the upper state rotational quantum number $J$ is 82.

37 parameters were varied. The resulting weighted dimensionless standard deviation of the fit is 1.05 and the *RMS* of the residuals is 0.00084 cm$^{-1}$. The ($\nu_{obs}$-$\nu_{EH}$) residuals together with measurement uncertainties are plotted in the bottom panel of Fig. 9.

**Table 10.** $^{17}O^{12}C^{18}O$ experimental data and summary of the Ritz and the EH line position fits.

| Reference | Sample | Isotopologues[1] | Setup[2] | Correction factor | $N_{fit}$[3] | $N_{excl}$[4] | $F_{\min}$ - $F_{\max}$ ($cm^{-1}$)[5] | $\Delta_{min}$ - $\Delta_{max}$[6] ($10^{-3}$ $cm^{-1}$) | $RMS_{\text{Ritz}}$ ($10^{-3}$ $cm^{-1}$) | $RMS_{\text{GIP}}$ ($10^{-3}$ $cm^{-1}$) |
|---|---|---|---|---|---|---|---|---|---|---|
| Claveau et al. [23,31] | $^{17}O$ | 4,8,9 | FTS | 1.0 | 682 | 6 | 571.1 - 2358.3 | 0.05 - 0.29 | 0.08 | 0.11 |
| Reisfeld et al. [30] | $^{17}O$ | 8,9 | FTS | 0. 9999941[a] | 65 | 13 | 628.4 - 690.4 | 3.0 | 2.95 | 2.95 |
| Lyulin et al. [13] (LADIR) | $^{17}O$ $^{18}O$ | 1-12 | FTS | 1.0 | 2816 | 260 | 1844.8 - 6908.4 | 0.1-1.0 | 0.41 | 0.63 |
| Toth et al. [32] | $^{18}O$ | 3,7,8 | FTS | 1.0 | 693 | 3 | 2258.0 - 5040.8 | 0.13 - 1.1 | 0.41 | 0.47 |
| Jacquemart et al. [33] | $^{17}O$ $^{18}O$ | 1,3,4,6-9 | FTS | 1.0 | 391 | 2 | 2268.7 - 5043.6 | 0.04 - 0.12 | 0.05 | 0.11 |
| Borkov et al. [25] | $^{17}O$ $^{18}O$ | 1-12 | FTS | 1.0 | 2726 | 0 | 3214.9 - 4681.3 | 0.8 | 0.67 | 1.24 |
| This work | $^{18}O$ | 1-11 | CRDS | 1.0 | 1661 | 10 | 5868.1 - 6908.4 | 1.0 | 0.62 | 1.29 |
| Lyulin et al. [13] (USTC) | $^{17}O$ $^{18}O$ | 4,8,9,12 | FTS | 0.99999996[a] | 404 | 0 | 5912.3 - 6908.5 | 0.7 | 0.40 | 0.54 |
| de Ghellinck d'Elseghem Vaernewijck et al. [16] | $^{17}O$ | 8.9 | CEAS | 0.9999988[a] | 217 | 4 | 6032.4 - 6411.6 | 2.0 | 2.53 | 2.77 |
| de Ghellinck d'Elseghem Vaernewijck et al. [15] | $^{17}O$ | 8,9 | CEAS | 0.9999993[a] | 32 | 8 | 6078.6 - 6100.1 | 1.7 | 1.57 | 1.71 |
| Golebiowski et al. [29] | $^{17}O$ | 4,8,9 | CEAS | 1.0 | 140 | 2 | 8035.7 - 8196.7 | 1.2 - 1.3 | 1.04 | 1.75 |
| number of fitted data | | | | | | | | | 9456 | 9754 |
| number of energy levels checked using the Ritz analysis / number of the EH parameters | | | | | | | | | 3563 | 63 |
| *RMS* of residuals ($10^{-3}$ $cm^{-1}$) | | | | | | | | | 0.70 | 1.08 |
| weighted dimensionless standard deviation $\chi$ | | | | | | | | | 1.08 | 1.40 |

Notes

[1] Isotopologues identified in the experimental spectra according to notation: 1: $^{12}C^{16}O_2$; 2: $^{13}C^{16}O_2$; 3: $^{16}O^{12}C^{18}O$; 4: $^{16}O^{12}C^{17}O$; 5: $^{16}O^{13}C^{18}O$; 6: $^{16}O^{13}C^{17}O$; 7: $^{12}C^{18}O_2$; 8: $^{17}O^{12}C^{18}O$; 9: $^{12}C^{17}O_2$; 10: $^{13}C^{18}O_2$; 11: $^{17}O^{13}C^{18}O$; 12: $^{13}C^{17}O_2$

[2] FTS - Fourier transform spectroscopy, CEAS - cavity-enhanced absorption spectroscopy, CRDS - cavity ring down spectroscopy

[3] $N_{fit}$ – number of lines of a given source included into the EH fit

[4] $N_{excl}$ – number of lines of a given source excluded from the EH fit

[5] $F_{\min}$ and $F_{\max}$ - minimum and maximum values of the wavenumbers in a given source

[6] $\Delta_{min}$ and $\Delta_{max}$ - minimum and maximum values of the measurement uncertainties for a given source

[a] correction factor was determined from Ritz check of the most abundant isotopologue of a sample

**Table 11.** $^{16}O^{13}C^{17}O$ experimental data and summary of the Ritz and the EH line position fits.

| Reference | Sample | Isotopologues[1] | Setup[2] | Correction factor | $N_{fit}$ [3] | $N_{excl}$ [4] | $F_{\min}$ - $F_{\max}$ ($cm^{-1}$)[5] | $\Delta_{min}$ - $\Delta_{max}$ [6] ($10^{-3}$ $cm^{-1}$) | $RMS_{\mathrm{Ritz}}$ ($10^{-3}$ $cm^{-1}$) | $RMS_{\mathrm{GIP}}$ ($10^{-3}$ $cm^{-1}$) |
|---|---|---|---|---|---|---|---|---|---|---|
| Teffo et al. [39] | $^{17}O$ | 6,11,12 | FTS | 1.0 | 99 | 2 | 617.5 - 2305.0 | 0.03 - 0.05 | 0.01 | 0.08 |
| Jolma [37] | $^{13}C$ | 2,3,5,6 | FTS | 0.99999992[a] | 63 | 1 | 618.2 - 675.0 | 0.19 - 0.24 | 0.25 | 0.32 |
| Lyulin et al. [13] (LADIR) | $^{17}O$ $^{18}O$ | 1-12 | FTS | 1.0 | 781 | 79 | 1977.0 - 4983.9 | 0.15 1.0 | 0.66 | 0.94 |
| Esplin et al. [38] | $^{13}C$ $^{18}O$ | 1,2,3,5,6,7 | FTS | 0.99999990[a] | 96 | 3 | 2203.9 - 2316.1 | 0.30 - 2.8 | 1.52 | 1.61 |
| Baldacci et al. [36] | natural | 2,5,6 | VGIS | 0.99999990[a] | 56 | 4 | 2229.2 - 2297.5 | 1.6 | 2.97 | 3.20 |
| Borkov et al. [25] | $^{17}O$ $^{18}O$ | 1-12 | FTS | 1.0 | 1036 | 0 | 3234.6 - 4680.1 | 0.8 | 0.53 | 0.86 |
| Toth et al. [41] | $^{18}O$ | 2,5,6,10 | FTS | 1.0 | 160 | 4 | 3476.4 - 4978.8 | 0.2 - 1.0 | 0.56 | 0.69 |
| Ding et al. [18] | $^{13}C$ | 2,5,6 | FTS | 0.99999998[a] | 418 | 5 | 4489.2 - 6750.9 | 0.53 - 1.1 | 0.74 | 1.06 |
| Mandin [35] | Venus | 1-7 | FTS | 0.99999938[a] | 46 | 6 | 4507.6 - 4736.4 | 1.2 - 2.5 | 2.57 | 3.02 |
| Perevalov et al. [17] | $^{13}C$ | 2,5,6,10,11 | CRDS | 1.0 | 627 | 9 | 5875.1 - 6745.0 | 0.6 - 2.1 | 1.18 | 1.72 |
| Perevalov et al. [19] | $^{13}C$ | 1-6 | CRDS | 1.0 | 231 | 19 | 6005.8 - 6768.6 | 0.7 - 1.5 | 0.89 | 1.65 |
| This work | $^{18}O$ | 1-11 | CRDS | 1.0 | 335 | 0 | 6044.5 - 6768.6 | 1.0 | 0.64 | 0.88 |
| Campargue et al. [42] | $^{13}C$ | 2,5,6,10,11 | CRDS | 1.0 | 310 | 22[b] | 7242.8 - 7917.3 | 0.61 - 0.80 | 0.65 | 1.47 |
| number of fitted data | | | | | | | | | 3927 | 4259 |
| number of energy levels checked using the Ritz analysis/ number of the EH parameters | | | | | | | | | 1819 | 43 |
| *RMS* of residuals ($10^{-3}$ $cm^{-1}$) | | | | | | | | | 0.89 | 1.17 |
| weighted dimensionless standard deviation $\chi$ | | | | | | | | | 1.43 | 1.36 |

Notes

[1] Isotopologues identified in the experimental spectra according to notation: 1: $^{12}C^{16}O_2$; 2: $^{13}C^{16}O_2$; 3: $^{16}O^{12}C^{18}O$; 4: $^{16}O^{12}C^{17}O$; 5: $^{16}O^{13}C^{18}O$; 6: $^{16}O^{13}C^{17}O$; 7: $^{12}C^{18}O_2$; 8: $^{17}O^{12}C^{18}O$; 9:

[2] FTS - Fourier transform spectroscopy, VGIS - vacuum grating infrared spectrograph, CRDS - cavity ring down spectroscopy

[3] $N_{fit}$ – number of lines of a given source included into the EH fit

[4] $N_{excl}$ – number of lines of a given source excluded from the EH fit

[5] $F_{\min}$ and $F_{\max}$ - minimum and maximum values of the wavenumbers in a given source

[6] $\Delta_{min}$ and $\Delta_{max}$ - minimum and maximum values of the measurement uncertainties for a given source

[a] correction factor was determined from Ritz check of the most abundant isotopologue of a sample

[b] Nineteen lines of the 40014-00001 band were excluded because of an interpolyad interaction of the 40014 upper state by the 60007 state

**Table 12.** $^{17}O^{13}C^{18}O$ experimental data and summary of the Ritz and the EH line position fits.

| Reference | Sample | Isotopologues[1] | Setup[2] | Correction factor | $N_{fit}$[3] | $F_{min}$ - $F_{max}$ (cm$^{-1}$)[4] | $\Delta_{min}$ - $\Delta_{max}$[5] (10$^{-3}$ cm$^{-1}$) | $RMS_{Ritz}$ (10$^{-3}$ cm$^{-1}$) | $RMS_{GIP}$ (10$^{-3}$ cm$^{-1}$) |
|---|---|---|---|---|---|---|---|---|---|
| Teffo et al. [39] | $^{17}O$ | 4,6,11,12 | FTS | 1.0 | 81 | 623.6 2281.0 | 0.03 - 0.20 | 0.033 | 0.11 |
| Lyulin et al. [13] (LADIR) | $^{17}O$ $^{18}O$ | 3-12 | FTS | 1.0 | 652 | 2182.0 4798.2 | 0.29 - 1.0 | 0.37 | 0.62 |
| Borkov et al. [25] | $^{17}O$ $^{18}O$ | 1-12 | FTS | 1.0 | 775 | 3379.3 4680.4 | 0.8 | 0.39 | 0.78 |
| This work | $^{18}O$ | 3-11 | CRDS | 1.0 | 253 | 5937.7 6714.1 | 1.0 | 0.71 | 1.18 |
| Perevalov et al. [17] | $^{13}C$ | 2 5 6 10 11 | CRDS | 1.00000003[a] | 129 | 5946.4 6714.2 | 0.9 - 1.4 | 0.94 | 1.37 |
| Campargue et al. [42] | $^{13}C$ | 2 5 6 10 11 | CRDS | 1.00000001[a] | 30 | 7837.4 7886.7 | 0.83 | 0.71 | 1.23 |
| number of fitted data | | | | | | | | 1623 | 1920 |
| number of energy levels checked using the Ritz analysis/ number of the EH parameters | | | | | | | | 874 | 37 |
| *RMS* of residuals (10$^{-3}$ cm$^{-1}$) | | | | | | | | 0.50 | 0.84 |
| weighted dimensionless standard deviation $\chi$ | | | | | | | | 0.97 | 1.05 |

Notes

[1] Isotopologues identified in the experimental spectra according to notation: 1: $^{12}C^{16}O_2$; 2: $^{13}C^{16}O_2$; 3: $^{16}O^{12}C^{18}O$; 4: $^{16}O^{12}C^{17}O$; 5: $^{16}O^{13}C^{18}O$; 6: $^{16}O^{13}C^{17}O$; 7: $^{12}C^{18}O_2$; 8: $^{17}O^{12}C^{18}O$; 9:

[2] FTS - Fourier transform spectroscopy, CRDS - cavity ring down spectroscopy

[3] $N_{fit}$ – number of lines of a given source included into the EH fit. All the lines were included.

[4] $F_{min}$ and $F_{max}$ - minimum and maximum values of the wavenumbers in a given source

[5] $\Delta_{min}$ and $\Delta_{max}$ - minimum and maximum values of the measurement uncertainties for a given source

[a] Correction factor was determined from Ritz check of the most abundant isotopologue of a sample.

## 6. Fitting of the line intensities

In this section, we use our measured intensities of the lines of $^{12}C^{17}O_2$, $^{16}O^{12}C^{17}O$, $^{17}O^{12}C^{18}O$, $^{16}O^{13}C^{17}O$ and $^{17}O^{13}C^{18}O$ to fit their effective dipole moment parameters. To our knowledge, no line intensity measurements in the 5850-7000 cm$^{-1}$ region were previously published for these isotopologues. Line intensity calculations used the eigenfuctions of the effective Hamiltonians derived in the preceding section for the $^{17}O^{12}C^{18}O$, $^{16}O^{13}C^{17}O$ and $^{17}O^{13}C^{18}O$ isotopologues and those for $^{12}C^{17}O_2$, $^{16}O^{12}C^{17}O$ published in our recent paper [25]. The used effective operator approach for the line intensity modeling has been presented in Refs. [6,44,45] and the computer code used to fit the EDM parameters is described in Ref. [46]. The total internal partition functions of $^{12}C^{17}O_2$, $^{16}O^{12}C^{17}O$, $^{17}O^{12}C^{18}O$ and $^{16}O^{13}C^{17}O$ were taken from Ref. [47] and that of $^{17}O^{13}C^{18}O$ from Ref. [48].

The fit minimizes the value of the dimensionless standard deviation, $\chi$, defined as

$$\chi = \sqrt{\frac{\sum_{i=1}^{N} \left( \left( S_i^{obs} - S_i^{calc} \right) / \delta_i \right)^2}{N - n}} , \qquad (2)$$

where $S_i^{obs}$ and $S_i^{calc}$ are, respectively, measured and calculated values of the intensity for the $i^{th}$ line, $\delta_i = (S_i^{obs} \sigma_i)/100\%$ is the absolute measurement error of the $i^{th}$ line and $\sigma_i$ the measurement error in %. $N$ and $n$ are the number of fitted line intensities and adjusted EDM parameters, respectively.

Another value used to characterize the quality of the fit is the root mean squares deviation:

$$RMS = \sqrt{\frac{\sum_{i=1}^{N} \left( \left( S_i^{obs} - S_i^{calc} \right) / S_i^{obs} \right)^2}{N}} \times 100\ \% . \qquad (3)$$

As a consequence of the weakness of the lines due to the very small abundances of the five $^{17}O$ isotopologues in our sample, a value of 10 % was used for the uncertainties of our relative line intensities. It is difficult to provide reliable error bars on the absolute intensity values as they were calculated on the basis of a statistical distribution of the O atoms between the various isotopologues. As no other experimental source has provided intensity measurements for the considered isotopologues in the region, the validity of this assumption cannot be tested and significant deviations may exist in particular in the present case where the five less abundant species are considered. Nevertheless, the fitted values obtained for the leading term of the EDM (see below) are fully consistent with that of the other isotopologues, indicating that our abundance values (Table 1) are reasonable.

The characteristics of the input data and the results of the line intensity fits are presented in Table 13. The fitted sets of the EDM parameters are given in Table 14. As it has been already mentioned in Section 3, the line intensities of the $^{12}C^{17}O_2$ bands assigned in the spectrum of the $^{17}O$ enriched water sample [21] were not determined and therefore the respective transitions are not involved into line intensity fit. The band by band intensity fit statistics is summarized in Table 15. The value of mean residual (*MR*) for a band presented in this table is defined by the following equation

$$MR = \frac{1}{N}\sum_{i=1}^{N}\left(S_i^{obs} - S_i^{calc}\right) / S_i^{obs} \times 100\% \ , \qquad (4)$$

where $N$ is the number of fitted line intensities for a given band.

**Table 13.** Experimental data and the results of the line intensity fits.

| Isotopologue | $\Delta P$ | $N_{band}$ [1] | $N_{fit}$ [2] | $N_{par}$ [3] | $\chi$ | *RMS* (%) |
|---|---|---|---|---|---|---|
| $^{16}O^{12}C^{17}O$ | 8 | 2 | 201 | 2 | 1.18 | 11.8 |
| | 9 | 18 | 1396 | 10 | 0.95 | 9.5 |
| $^{17}O^{12}C^{18}O$ | 8 | 1 | 53 | 1 | 0.74 | 7.3 |
| | 9 | 18 | 1446 | 10 | 0.82 | 8.2 |
| $^{16}O^{13}C^{17}O$ | 9 | 5 | 268 | 4 | 0.80 | 7.9 |
| $^{17}O^{13}C^{18}O$ | 9 | 4 | 230 | 4 | 0.99 | 9.9 |
| $^{12}C^{17}O_2$ | 9 | 3 | 164 | 4 | 1.22 | 12.1 |

Notes
[1] $N_{band}$ – number of bands included in the fit
[2] $N_{fit}$ – number of lines included in the fit
[3] $N_{par}$ – number of adjusted parameters.

**Table 14.** Effective dipole moment parameters for the $\Delta P$= 8 and $\Delta P$= 9 series of transitions of $^{16}O^{12}C^{17}O$, $^{17}O^{12}C^{18}O$, $^{16}O^{13}C^{17}O$, $^{17}O^{13}C^{18}O$ and $^{12}C^{17}O_2$

| Parameter[a] | $\Delta V_1$ | $\Delta V_2$ | $\Delta V_3$ | $\Delta \ell_2$ | Value[b] | | | | | Order |
|---|---|---|---|---|---|---|---|---|---|---|
| | | | | | $^{16}O^{12}C^{17}O$ | $^{17}O^{12}C^{18}O$ | $^{16}O^{13}C^{17}O$ | $^{17}O^{13}C^{18}O$ | $^{12}C^{17}O_2$ | |
| $\Delta P$= 8 | | | | | | | | | | |
| $M$ | 1 | 0 | 2 | 0 | -0.10435(38) | -0.10789(54) | | | | $10^{-3}$ |
| $M$ | 0 | 2 | 2 | 0 | 0.434(19) | | | | | $10^{-5}$ |
| $\Delta P$= 9 | | | | | | | | | | |
| $M$ | 3 | 0 | 1 | 0 | -0.25995(35) | -0.26652(41) | -0.23389(10) | -0.2546(16) | -0.2539(19) | $10^{-3}$ |
| $b_J$ | 3 | 0 | 1 | 0 | 0.307(42) | | 0.55(19) | 1.19(26) | 0.505(24) | $10^{-3}$ |
| $M$ | 2 | 2 | 1 | 0 | 0.2351(19) | 0.2463(22) | 0.2190(75) | 0.284(20) | 0.245(18) | $10^{-4}$ |
| $M$ | 1 | 4 | 1 | 0 | -0.1177(57) | -0.0931(67) | | | | $10^{-5}$ |
| $M$ | 0 | 6 | 1 | 0 | 0.379(92) | 0.721(93) | | | | $10^{-7}$ |
| $M$ | 0 | 0 | 3 | 0 | 0.30688(73) | 0.30398(10) | 0.29755(92) | 0.3105(12) | 0.3002(25) | $10^{-3}$ |
| $\kappa_1$ | 0 | 0 | 3 | 0 | | -0.146(17) | | | | |
| $\kappa_2$ | 0 | 0 | 3 | 0 | | 0.107(31) | | | | $10^{-1}$ |
| $M$ | 1 | 1 | 2 | 1 | 0.21301(81) | 0.22355(76) | | | | $10^{-4}$ |
| $b_J$ | 1 | 1 | 2 | 1 | 0.317(23) | 0.269(13) | | | | $10^{-2}$ |
| $M$ | 0 | 3 | 2 | 1 | -0.983(30) | -1.090(26) | | | | $10^{-6}$ |
| $b_J$ | 0 | 3 | 2 | 1 | 0.875 (18) | | | | | $10^{-2}$ |

[a] The parameters $M$ are given in Debye while the $\kappa_i$ and $b_J$ parameters are dimensionless. Only relative signs of the $M$ parameters within a given series of transitions are determined.

[b] The numbers in parentheses correspond to one standard deviation in units of the last quoted digit.

Table 15. Band by band statistics of the intensity fits.

| Band | Σ [a] cm/molecule | *N* [b] | $J_{min}$ $J_{max}$ [c] | $\nu_{min}$ $\nu_{max}$ [d] ($cm^{-1}$) | $S_{min}$ $S_{max}$ [e] $10^{-28}$ cm/molecule | *MR*(%) | *RMS*(%) | *MaxR*(%) [f] |
|---|---|---|---|---|---|---|---|---|
| 627 Δ*P* = 8 | | | | | | | | |
| 10022 00001 | 3.56e-25 | 68 | 1 53 | 5857.7 5908.9 | 2.79-100 | -0.3 | 12.7 | 29.1 |
| 10021 00001 | 3.76e-25 | 67 | 0 48 | 5936.6 6008.7 | 4.72-120 | -0.6 | 9.3 | 24.9 |
| 627 Δ*P* = 9 | | | | | | | | |
| 31114 01101 | 5.36e-26 | 62 | 4 46 | 5937.9 6003.4 | 1.12-14.4 | -1.5 | 14.1 | 26.8 |
| 30014 00001 | 1.14e-24 | 98 | 0 61 | 5980.8 6069.6 | 1.94-230 | -1.2 | 8.2 | 27.6 |
| 31113 01101 | 4.60e-25 | 129 | 3 56 | 6092.4 6176.8 | 1.13-74.9 | -2.2 | 9.9 | 23.5 |
| 40014 10002 | 7.00e-27 | 12 | 3 37 | 6093.1 6146.2 | 2.36-9.38 | -5.1 | 10.2 | 18.0 |
| 40013 10001 | 8.08e-27 | 23 | 3 40 | 6113.7 6172.8 | 1.09-5.47 | -5.9 | 14.9 | 28.9 |
| 30013 00001 | 8.52e-24 | 102 | 0 68 | 6118.7 6209.7 | 3.14-1840 | -2.1 | 6.6 | 22.6 |
| 40013 10002 | 1.26e-26 | 29 | 1 42 | 6214.4 6273.2 | 0.966-7.37 | -8.3 | 16.0 | 29.1 |
| 30012 00001 | 6.53e-24 | 122 | 0 74 | 6235.9 6336.8 | 0.312-1280 | -1.1 | 7.2 | 27.1 |
| 31112 01101 | 4.21e-25 | 172 | 1 60 | 6254.8 6342.1 | 0.446-54.5 | -2.8 | 10.3 | 29.0 |
| 40012 10001 | 9.56e-27 | 37 | 2 46 | 6265.7 6329.2 | 0.468-4.52 | 1.5 | 14.2 | 29.7 |
| 30011 00001 | 5.78e-25 | 103 | 0 65 | 6409.5 6505.7 | 0.244-132 | 0.1 | 8.5 | 25.1 |
| 31111 01101 | 4.73e-26 | 97 | 4 50 | 6454.5 6527.0 | 0.367-9.55 | -1.8 | 12.3 | 27.3 |
| 11122 00001 | 3.34e-26 | 93 | 0 51 | 6464.3 6528.7 | 0.182-7.71 | 0.0 | 10.7 | 29.1 |
| 40011 10001 | 2.80e-28 | 6 | 16 30 | 6508.8 6517.8 | 0.268-0.55 | -4.4 | 11.5 | 19.2 |
| 11121 00001 | 3.14e-26 | 70 | 1 48 | 6621.5 6669.4 | 0.411-10.6 | 0.0 | 12.8 | 29.8 |
| 01131 01101 | 1.64e-24 | 133 | 1 60 | 6831.4 6924.1 | 1.74-208 | 0.7 | 7.4 | 26.2 |
| 02231 02201 | 6.26e-27 | 10 | 19 31 | 6839.4 6852.6 | 4.42-8.83 | 6.2 | 15.1 | 22.0 |
| 00031 00001 | 2.45e-23 | 98 | 0 63 | 6866.8 6961.2 | 28.1-5170 | -2.9 | 4.9 | 14.9 |
| 637 Δ*P* = 9 | | | | | | | | |
| 30013 00001 | 1.93e-26 | 28 | 3 33 | 6044.5 6093.3 | 3.55-9.47 | -0.9 | 9.2 | 19.7 |
| 30012 00001 | 4.83e-26 | 43 | 2 43 | 6150.5 6213.1 | 2.54-17.2 | 1.3 | 8.8 | 16.8 |
| 30011 00001 | 5.32e-27 | 31 | 2 40 | 6296.1 6344.4 | 0.542-3.12 | -1.0 | 9.4 | 19.1 |
| 01131 01101 | 1.26e-26 | 80 | 1 39 | 6675.4 6733.4 | 0.325-2.41 | -0.5 | 8.9 | 20.4 |
| 00031 00001 | 2.45e-25 | 86 | 0 55 | 6685.3 6768.6 | 0.985-55.4 | 0.5 | 4.7 | 18.1 |
| 728 Δ*P* = 8 | | | | | | | | |
| 10021 00001 | 2.19e-25 | 53 | 1 52 | 5894.8 5950.9 | 10.3- 70.4 | 0.0 | 7.3 | 16.9 |
| 728 Δ*P* = 9 | | | | | | | | |
| 31114 01101 | 1.43e-25 | 88 | 1 45 | 5868.1 5915.7 | 2.57-27.1 | -2.1 | 9.2 | 26.2 |
| 30014 00001 | 2.63e-24 | 103 | 0 65 | 5891.0 5981.3 | 2.53-530 | 2.6 | 6.8 | 20.5 |
| 40014 10002 | 4.81e-26 | 41 | 3 44 | 5989.0 6047.1 | 2.25-17.7 | -1.4 | 9.7 | 24.5 |
| 31113 01101 | 6.52e-25 | 140 | 1 60 | 6004.7 6078.9 | 1.11-82 | 0.0 | 7.4 | 24.4 |
| 30013 00001 | 1.17e-23 | 121 | 0 68 | 6012.4 6105.1 | 5.12-2310 | 0.0 | 5.5 | 23.4 |
| 40013 10001 | 5.80e-27 | 13 | 4 34 | 6033.2 6063.6 | 2.68-6.08 | -2.8 | 10.1 | 17.1 |
| 40013 10002 | 8.36e-27 | 20 | 4 41 | 6139.5 6178.8 | 1.26-6.70 | -5.4 | 11.5 | 23.5 |
| 30012 00001 | 3.89e-24 | 101 | 0 68 | 6148.6 6244.7 | 1.12-916 | 0.7 | 6.4 | 23.9 |
| 31112 01101 | 3.31e-25 | 129 | 1 53 | 6175.7 6250.2 | 1.56-50.3 | 0.9 | 8.6 | 22.8 |
| 40012 10001 | 3.59e-27 | 18 | 1 39 | 6189.9 6239.3 | 0.544-3.34 | -5.7 | 10.3 | 26.1 |
| 30011 00001 | 3.26e-25 | 86 | 0 51 | 6352.8 6424.1 | 2.69-68.8 | 0.7 | 6.2 | 17.3 |
| 31111 01101 | 2.83e-26 | 71 | 4 37 | 6396.2 6446.8 | 1.39-6.58 | -0.4 | 13.3 | 27.5 |
| 11122 00001 | 3.75e-26 | 78 | 1 54 | 6409.7 6463.8 | 0.429-10 | -0.5 | 8.6 | 20.6 |
| 11121 00001 | 3.26e-26 | 74 | 1 56 | 6555.2 6608.4 | 0.179-9.94 | 0.7 | 8.3 | 23.3 |
| 01131 01101 | 1.70e-24 | 141 | 1 67 | 6772.0 6871.7 | 0.499-241 | 1.2 | 7.6 | 21.5 |
| 02231 02201 | 4.21e-26 | 70 | 2 39 | 6779.9 6834.8 | 1.76-10.4 | -0.8 | 11.3 | 22.5 |
| 10031 10001 | 7.89e-27 | 17 | 5 41 | 6782.3 6838.1 | 1.52-7.24 | -2.6 | 9.6 | 21.5 |
| 10032 10002 | 1.29e-26 | 17 | 7 35 | 6804.5 6848.8 | 4.04-12.1 | 1.0 | 7.2 | 21.4 |
| 00031 00001 | 2.82e-23 | 118 | 0 69 | 6805.1 6908.4 | 8.32-5630 | -0.9 | 5.5 | 22.8 |
| 727 Δ*P* = 9 | | | | | | | | |
| 30013 00001 | 1.37e-25 | 54 | 0 43 | 6084.5 6146.8 | 3.43-47.1 | 0.0 | 13.0 | 26.1 |
| 30012 00001 | 7.44e-26 | 57 | 1 52 | 6204.5 6276.7 | 0.553-29.3 | 0.0 | 12.5 | 34.6 |
| 00031 00001 | 3.92e-25 | 53 | 0 41 | 6873.5 6932.8 | 10.2-149 | 0.0 | 10.3 | 26.3 |
| 738 Δ*P* = 9 | | | | | | | | |
| 30013 00001 | 4.34e-26 | 39 | 3 41 | 5940.7 5998.5 | 2.55-18.4 | 0.0 | 10.3 | 24.0 |
| 30012 00001 | 3.72e-26 | 34 | 4 39 | 6062.7 6111.5 | 3.77-16.1 | 0.0 | 13.2 | 25.4 |
| 01131 01101 | 1.06e-26 | 66 | 1 45 | 6616.2 6678.1 | 0.221-2.65 | -2.4 | 10.9 | 22.0 |
| 00031 00001 | 2.59e-25 | 91 | 0 63 | 6625.8 6714.1 | 0.343-56.4 | 1.8 | 6.9 | 23.7 |

*Notes*

[a] Σ – sum of the measured line intensities for a given band

[b] *N* – number of the measured line intensities for a given band

[c] $J_{min}$ $J_{max}$ – minimum and maximum values of the angular momentum quantum number for a given band

[d] $\nu_{min}$ $\nu_{max}$ – minimum and maximum values of the line positions for a given band

[e] $S_{min}$ $S_{max}$ – minimum and maximum values of the line intensities for a given band

[f] *MaxR* – maximum residual between observed and calculated values of the line intensities for a given band.

## 7. Discussion and conclusion

This contribution concludes our analysis and theoretical modeling of the absorption spectrum of highly $^{18}$O enriched carbon dioxide recorded by very high sensitivity CW-Cavity Ring Down spectroscopy between 5851 and 6990 cm$^{-1}$ (1.71-1.43 µm) [1]. In this third report, we have presented the analysis of the spectrum of the five $^{17}$OCO isotopologues, which are the least abundant in our sample but significantly enriched compared to natural carbon dioxide (relative abundances of about 2% for $^{16}O^{12}C^{17}O$ and $^{17}O^{12}C^{18}O$ and less than $4\times10^{-4}$ for $^{16}O^{13}C^{17}O$, $^{17}O^{13}C^{18}O$ and $^{12}C^{17}O_2$). As a result of the very high sensitivity of the recordings, 4704 lines of 64 bands could be assigned; most of them are newly reported.

Overall, about 16000 $CO_2$ lines have been assigned in the present work and in the two previous. Taking into account 3587 lines assigned to $^{12}C^{16}O_2$ and $^{13}C^{16}O_2$ and about 1350 impurity lines assigned to $CH_4$, HCN, CO and $H_2O$ isotopologues, it leads to a total number of about 20900 lines identified, leaving only about 947 unassigned lines (about 4% of the total). This result relies mainly on the very high predictive capability of the effective models developed for each of the studied isotopologues. All the unassigned lines are very weak lines not organised in regular series. They are provided as Supplementary Material for all intents and purposes.

The newly measured line positions (estimated accuracy between $6\times10^{-4}$ and $1\times10^{-3}$ cm$^{-1}$) allowed us improving the sets of the effective Hamiltonian parameters for nine isotopologues. The parameters for $^{17}O^{12}C^{18}O$, $^{16}O^{13}C^{17}O$ and $^{17}O^{13}C^{18}O$ are presented in this paper and those for $^{16}O^{12}C^{18}O$, $^{12}C^{18}O_2$, $^{13}C^{18}O_2$, $^{16}O^{13}C^{18}O$, $^{16}O^{12}C^{17}O$ and $^{12}C^{17}O_2$ were published in Refs. [1,2,25]. The quality of the line position calculations achieved with the refined sets of the effective Hamiltonian parameters is illustrated in Fig.10 which includes a few bands assigned after the refinement of the EH parameters. The deviations which are, for these bands, the differences between the predicted and measured line positions are found at the same level than for those of the fitted lines.

From the set of measured line intensities, the effective dipole moment parameters for the $\Delta P$=8 and $\Delta P$=9 series of transitions in $^{16}O^{12}C^{17}O$, $^{17}O^{12}C^{18}O$, $^{16}O^{13}C^{17}O$, $^{17}O^{13}C^{18}O$ and $^{12}C^{17}O_2$ could be fitted for the first time. The (low) abundance values of these isotopologues in our sample were estimated on the basis of the stated oxygen atomic relative abundance, assuming a statistical distribution of the oxygen atoms within $CO_2$ species and a natural value for the carbon relative abundance. This approach seems to be validated (within an estimated uncertainty between 10% and 15%) by the abundance values obtained for the $^{12}C^{16}O_2$ and $^{13}C^{16}O_2$ species (see Section 2) and by the comparison of the obtained values of the fitted

EDM parameters for different isotopologues (see Table 14). Indeed, according to our theoretical study [49], the values of the principal vibrational EDM parameters for the series of transitions which are allowed for both symmetric and asymmetric isotopologues do not depend strongly on isotopic substitution of both oxygen and carbon atoms. For example, the values of the $M_{003}$ and $M_{301}$ parameters included in Table 14 for are very close to those of the principal isotopologue ($M_{003}$= 0.31068(34) ×$10^{-3}$ and $M_{301}$= -0.26621(13) ×$10^{-3}$ Debye [50]).

Part of the results obtained from the analysis of the present CRDS spectra has already been used to improve the carbon dioxide line list in the last edition of the HITRAN database [51]. A major improvement has been the inclusion of the $\Delta P$= 8 series of transitions of $^{16}O^{12}C^{18}O$ measured in our first report [1] which fall in a transparency window and dominate the spectrum in spite of the very low abundance of $^{16}O^{12}C^{18}O$ in natural carbon dioxide. The obtained parameters of the effective operators will be used for the generation of a new version of the CDSD data bank.

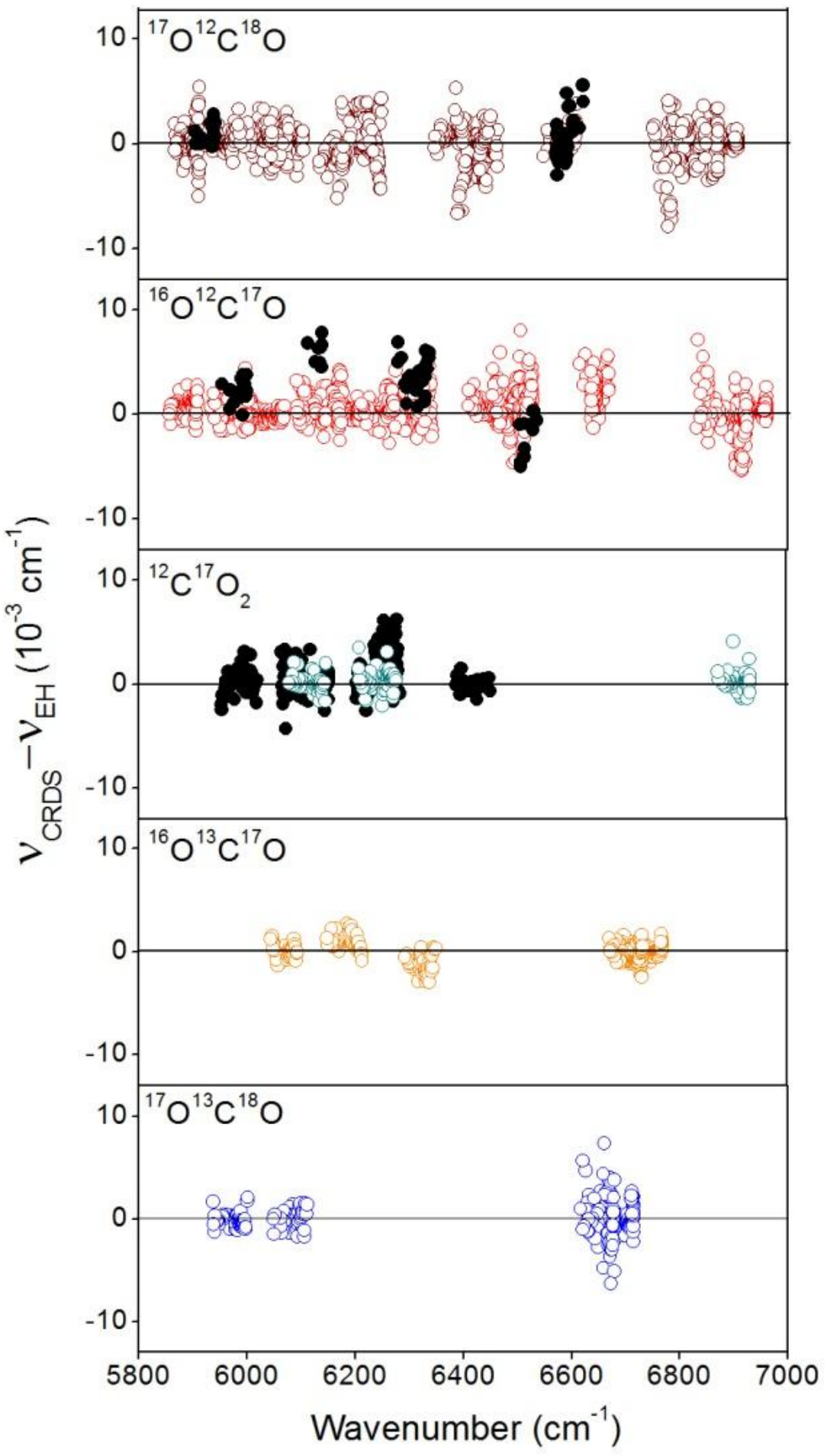

**Fig. 10.** Differences between the line positions of $^{17}O^{12}C^{18}O$, $^{16}O^{12}C^{17}O$, $^{12}C^{17}O_2$, $^{16}O^{13}C^{17}O$ and $^{17}O^{13}C^{18}O$ measured by CW-CRDS between 5850 and 7000 cm$^{-1}$ and calculated with the effective Hamiltonians. The black circles denote the few bands not used in the input data of the EH fit for $^{17}O^{12}C^{18}O$ (11121-01101, 12221-01101, 20021-01101), $^{16}O^{12}C^{17}O$ (11121-01101, 32211-02201, 32212-02201, 32213-02201) and $^{12}C^{17}O_2$ (30011-00001; 30014-00001, 31112-01101, 31113-01101 [21]).

**Acknowledgements**

The support of the Laboratoire International Associé SAMIA between CNRS (France) and RAS (Russia, RFBR grant N 12-05-93106) is acknowledged. E. V. Karlovets thanks the Embassy of France in Moscow for providing her with PhD grant.